\documentclass[a4paper,11pt]{article}

\usepackage{jheppub} 

\usepackage[T1]{fontenc} 
\usepackage{amsmath,amsthm}
\usepackage{amssymb,amsfonts}
\usepackage{bbm}
\usepackage{graphicx}
\usepackage{dsfont}
\usepackage{relsize}
\usepackage{stmaryrd}
\usepackage{lmodern}
\usepackage{slantsc}
\usepackage{scalefnt}
\usepackage{subfigure}
\usepackage{xspace}
\usepackage{boxedminipage}
\usepackage{ifpdf}
\usepackage{xifthen}
\usepackage{tensor}
\usepackage{array}
\usepackage{tabu}
\usepackage{pbox}
\usepackage{multirow}
\usepackage{tikz}
\usepackage{color}
\usepackage{diagbox}
\usetikzlibrary{arrows,matrix,calc,scopes,decorations.markings}
\allowdisplaybreaks[1]
\usepackage[mathcal]{euscript}

\newtheorem*{ansatz*}{Ansatz}

\newcommand{\be}{\begin{equation}}
\newcommand{\ee}{\end{equation}}
\newcommand{\bse}{\begin{subequations}}
\newcommand{\ese}{\end{subequations}}
\newcommand{\ket}[1]{\left|{#1}\right\rangle}
\newcommand{\bket}[1]{\Biggl|{#1}\Biggr\rangle}

\newcommand{\Z}{\mathbb{Z}}

\newcommand{\F}{\mathfrak{F}}

\newcommand{\T}{\mathcal{T}}

\newcommand{\bpm}{\begin{pmatrix}}
\newcommand{\epm}{\end{pmatrix}}
\newcommand{\bmm}{\begin{matrix}}
\newcommand{\emm}{\end{matrix}}

\newcommand{\Bb}{\overline{B}}
\newcommand{\x}{\times}

\newcommand{\rep}{\mathrm{Rep}}

\newcommand\numberthis{\addtocounter{equation}{1}\tag{\theequation}}

\tikzset{snake it/.style={decorate, decoration={snake,amplitude=0.15mm,segment length=1mm}}}
\tikzset{->-/.style={decoration={
                        markings,
                        mark=at position .55 with {\arrow{latex}}},postaction={decorate}}}

\newtabulinestyle{dash=off 2pt}

 {\null\,\vcenter\bgroup\scriptsize
  \arraycolsep=.13885em
  \hbox\bgroup$\array{@{}#1@{}}}
 {\endarray$\egroup\egroup\,\null}

\input{mtikz.tex}

\newcolumntype{C}{>{\centering\arraybackslash} m{1.5em} }

\makeatletter
\newcommand*{\Relbarfill@}{\arrowfill@\Relbar\Relbar\Relbar}
\newcommand*{\xeq}[2][]{\ext@arrow 0055\Relbarfill@{#1}{#2}}
\makeatother
\allowdisplaybreaks

\title{Extend The Levin-Wen Model To Two-dimensional Topological Orders With Gapped Boundary Junctions}
\date{\today}
\author[a,b]{Hongyu Wang}
\author[c,1]{Yuting Hu}
\author[a,b,d,1]{Yidun Wan\note{Corresponding author}}
\affiliation[a]{State Key Laboratory of Surface Physics,Department of Physics, Center for Field Theory and Particle Physics, and Institute for Nanoelectronic devices and Quantum computing, Fudan University, 
\\Shanghai 200433, China}
\affiliation[b]{Shanghai Qi Zhi Institute, 
\\Shanghai 200030, China}
\affiliation[c]{School of Physics, Hangzhou Normal University,
\\Hangzhou 311121, China}
\affiliation[d]{Zhangjiang Fudan International Innovation Center, Fudan University, 
\\Shanghai 201210, China}
\emailAdd{wanghy17@fudan.edu.cn, yuting.phys@gmail.com, ydwan@fudan.edu.cn}

\abstract{A realistic material may possess defects, which often bring the material new properties that have practical applications. The boundary defects of a two-dimensional topologically ordered system are thought of as an alternative way of realizing topological quantum computation. To facilitate the study of such boundary defects, in this paper, we construct an exactly solvable Hamiltonian model of topological orders with gapped boundary junctions, where the boundary defects reside, by placing the Levin-Wen model on a disk, whose gapped boundary is separated into multiple segments by junctions. We derive a formula of the ground state degeneracy and an explicit ground-state basis of our model. We propose the notion of mobile and immobile charges on the boundary and find that they are quantum observables and label the ground-state basis. Our model is computation friendly. }

\begin{document}

\maketitle

\flushbottom
\section{Introduction}\label{sec:intro}
 Recent studies \cite{Beigi2011,Bullivant2017,Kitaev2012,Kapustin2013,HungWan2014,Lan2014,HungWan2015a,Hu2017a,Hu2017,Chen2018,Li2019a,Wang2020} have revealed various novel properties of two-dimensional topologically ordered matter phases (topological orders) with gapped boundaries. More recently, it is shown that the gapped boundaries of certain topological orders may offer a new way of realizing topological quantum computation (TQC) \cite{Cong2016a}. In addition to usual boundary excitations, gapped boundaries of a topologically order system may also be habitable for boundary defects such as Majorana zero modes and parafermion zero modes \cite{Cheng2012,Lindner2012,Clarke2013,Barkeshli2013d,Sarma2015}. Such boundary defects may have non-Abelian braiding statistics \cite{Barkeshli2013b} and thus may also help realize TQC. Boundary defects are also related to topological defects in conformal field theory \cite{Frohlich2007}.

\begin{figure}[h!]
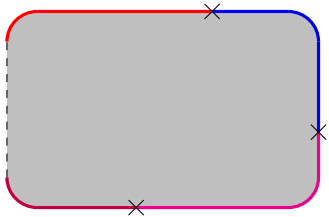

\centering
\disk
\caption{A disk with multiple boundary segments. Different colours represent different boundary segments. A cross represents a boundary junction between two adjacent boundary segments.}
\label{fig:disk}
\end{figure}

Physically, boundary defects can be realized at gapped boundary junctions \cite{Fu2008,Kitaev2012,Bridgeman2019,Bridgeman2020,Lou2021}. A gapped boundary junction joins two adjacent gapped boundary segments, as a physical boundary of a system may be divided into a few segments, on which there are different gapped boundary conditions (see Fig. \ref{fig:disk}). Nevertheless, boundary junctions may not be gapped in general. It is therefore natural to ask under what conditions boundary junctions are gapped. Answering this question is helpful for realizing topologically ordered systems with boundary defects in lab.      

In this paper, we address this question by constructing an exactly solvable Hamiltonian model with gapped boundary junctions based on the extended Levin-Wen (ELW) model \cite{Hu2017a}. The Levin-Wen model is a large class of Hamiltonian models of gapped topological orders on closed two-surfaces, with the input data being fusion categories. The ELW model extends the Levin-Wen model to the case of open surfaces. In the ELW model, a gapped boundary is characterized by a Frobenius algebra \cite{Kock2003,Fukuma1994,Fuchs2002,Fuchs2004a,Kirillov2002} object in the input fusion category. We generalize the ELW model to the case with multiple boundary segments characterized by distinct Frobenius algebras. A gapped boundary junction between two adjacent segments is characterized by either a morphism between or a common Frobenius subalgebra of the two Frobenius algebras characterizing the two segments. Our model leads to the following results.
\begin{enumerate}
    
    \item  A formula of the ground state degeneracy (GSD) and explicit ground-state basis on a disk with $N$ gapped boundary junctions.   %
    \item We propose the notion of \textit{mobile} and \textit{immobile charges} (defined in Section \ref{sec:grdA}) on the boundary. In the language of boundary condensation, an immobile charge is the total condensate charge residing in a boundary segment. We find that immobile charges are the quantum observables and label the ground-state basis.
\end{enumerate}

Our boundary junction Hamiltonian is constructed solely in terms of the input degrees of freedom (i.e., the simple objects in the input fusion category) of our model. Our approach is computation friendly, as can be seen from the concrete examples we provide. In Ref.\cite{Cong2016a, Cong2017a}, the authors constructed an exactly solvable quantum double model with gapped boundary junctions, which is a special case of our construction.

We organize our paper as follows. Section \ref{sec:torc} introduces the concept of junctions with a simple example---the ELW $\Z_2$ model. Section \ref{sec:ham} reviews the ELW model and constructs our model of gapped boundary junctions. Section \ref{sec:top} derives the GSD formula and the ground-state basis of our model. Section \ref{sec:exams} presents two explicit examples. Section \ref{sec:con} briefly discusses the elementary excitations on gapped boundary junctions. The Appendices collect a few technical calculations.

\section{The extended Levin-Wen $\Z_2$ model with gapped boundary junctions}\label{sec:torc}
 
To acquire certain intuition for systematically constructing our model, let us first consider a simple example: the ELW $\Z_2$ model with gapped boundary junctions. We shall write down the Hamiltonian of the ELW $\Z_2$ model with gapped boundary junctions and explicitly derive the ground-state wavefunctions. We show that the ground states are  degenerate in general and are characterized by configurations of the condensate charges on the boundary. 

\begin{figure}[h!]
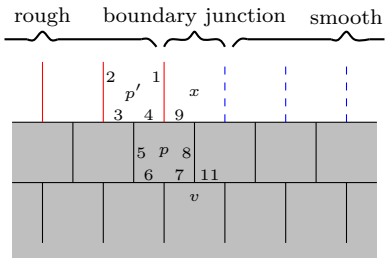

\centering
\toric
\caption{A portion of the trivalent lattice with a gapped boundary junction between a rough boundary segment and a smooth boundary segment. The grey region represents the bulk. Red (blue dashed) lines represent the dangling edges on rough (smooth) boundaries. The spins on the blue dashed lines are fixed to $-1$. $p$ and $p'$ label the bulk plaquettes and the boundary open plaquettes.  $v$ label the vertices. The boundary junction is the open plaquette labeled by $x$.}
\label{fig:tori}
\end{figure}

For simplicity, we define ELW $\Z_2$ model with gapped boundary junctions on a trivalent lattice (see Fig. \ref{fig:tori}). Residing on each edge of the lattice is a degree of freedom (also called a spin) taking value in $\Z_2$. In the literature, there are two kinds of gapped boundary segments: rough and smooth\cite{Bravyi1998,  Bombin2011,HungWan2015}. A gapped boundary junction is defined as the open boundary plaquette between the rough and smooth boundary segments. The Hilbert space is spanned by all possible configurations of the spins on the edges.  

The Hamiltonian of the ELW $\Z_2$ model with gapped boundary junctions reads
\begin{equation}
    H^{\Z_2}=H^{\Z_2}_{\text{(bulk)}}+H^{\Z_2}_{\text{(rough bdry)}},
\end{equation}
where
\begin{equation}
    H^{\Z_2}_{\text{(bulk)}}=-\sum_{v} Q^{\Z_2}_v-\sum_{p} B^{\Z_2}_p,
\end{equation}
\begin{equation}
    H^{\Z_2}_{\text{(rough bdry)}}=-\sum_{p'}\overline B^{\Z_2}_{p'},
\end{equation}
Here, the operators $Q^{\Z_2}_v$ are defined at the vertices $v$, the operators $B^{\Z_2}_p$ are defined on the bulk plaquettes $p$, and $\overline B^{\Z_2}_{p'}$ are defined on the rough boundary plaquettes $p'$. These operators are expressed in terms of the Pauli matrices. For example, the three operators respectively on $v$, $p$, and $p'$ in Fig. \ref{fig:tori} are
\be\begin{aligned}\label{eq:toriop}
&Q^{\Z_2}_v=\sigma_7^z\sigma_{11}^z\sigma_8^z,
\\
&B^{\Z_2}_p=\sigma_4^x\sigma_5^x\sigma_6^x\sigma_7^x\sigma_8^x\sigma_9^x,
\\
&\overline B^{\Z_2}_{p'}=\sigma_1^x\sigma_2^x\sigma_{3}^x\sigma_{4}^x.
\end{aligned}\ee
All the operators $Q^{\Z_2}_v$, $B^{\Z_2}_p$, and $\overline B^{\Z_2}_{p'}$ in the Hamiltonian have eigenvalues $\pm1$ and  commute with each other. Therefore, the model is exactly solvable. 

The energy eigenstates are common eigenstates of $Q^{\Z_2}_v$, $B^{\Z_2}_p$, and $\overline B^{\Z_2}_{p'}$ for all $v$, $p$, and $p'$. The ground states are the common $+1$ eigenstates, while an excitated state is a $-1$ eigenstate of one or more of these operators. 

In the bulk, there are two types of elementary excitations: If $Q^{\Z_2}_v=-1$ ($B^{\Z_2}_p=-1$) for some vertex $v$ (plaquette $p$), then there is a bulk charge excitation (bulk flux excitation). 

On the rough boundary, if $\Bb^{\Z_2}_{p'}=-1$ for some boundary plaquette $p'$, then there is a boundary flux excitation.

A bulk charge can disappear at the rough boundary segment. Such a phenomenon is called boundary charge condensation\cite{Beigi2011, Hung2013, Hu2021,Williamsona}. On the smooth boundary, however, a bulk charge cannot disappear and is identified with a boundary charge excitation\cite{Beigi2011,Bravyi1998}. 
     
\begin{figure}[h!]
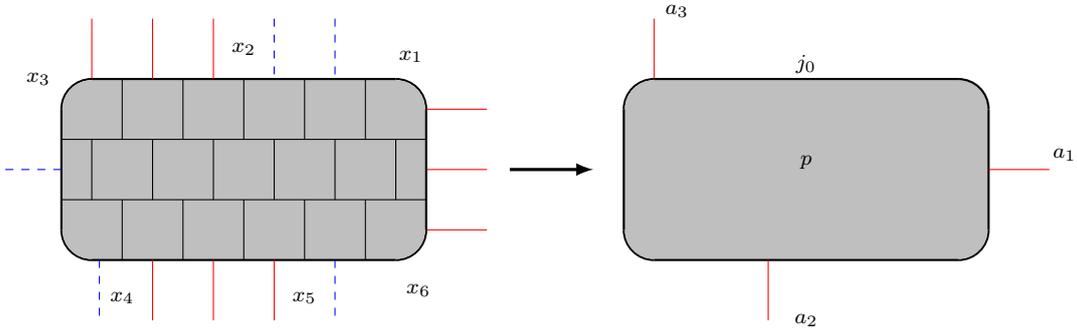

\centering
\Shrinkgraph
\caption{An example of the simplification of the lattice on a disk. The $6$ boundary junctions are labeled by $x_1$ to $x_6$ respectively. Red (blue dashed) lines represent the dangling edges on rough (smooth) boundary segments. The generalization to $2N$ boundary junctions is straightforward.}
\label{fig:shrinkgraph}
\end{figure}

If we consider the states without any bulk or boundary excitations, the lattice can be simplified using topology preserving mutations of the lattice\cite{Hu2017a} (see also Appendix \ref{sec:pach}). Let us consider the ELW $\Z_2$ model on a disk with $2N$ boundary junctions. The topology preserving mutations allow us to merge all the bulk plaquettes into a single bulk plaquette and combine all the dangling edges of each rough boundary segment into a single dangling edge (see Fig. \ref{fig:shrinkgraph}). The resultant simplified model is equivalent to an effective spin chain model, whose Hilbert space is spanned by $N+1$ spins $\mathcal{H_{\rm{sim}}}=\text{span} \{\ket{ j_0, a_1, a_2,\dots, a_{N-1}, a_{N}}\}$ (see Fig. \ref{fig:shrinkgraph}).  

On the simplified lattice, the effective spin chain Hamiltonian is
\be\label{eq:spinHam}
H^{\Z_2}_{\rm eff}=-\sigma_{j_0}^x-\prod_{i=1}^{N}\sigma_{a_{i}}^z,
\ee
where the first term descends from the bulk plaquette operator, and the second term is the global charge constraint that restricts the global charge to zero. 

We have two observations. First, $\sigma_{a_i}^z$ commutes with $H^{\Z_2}_{\rm eff}$ and hence is a conserved observable. Second, $\prod_{j=1}^N\sigma_{a_i}^z=1$ in the ground states. Hence, the ground states have a basis:
\be\label{eq:TCgdbasis}
\left\{\ket{\sigma_{j_0}^x=1, \sigma_{a_1}^z=\pm 1, \sigma_{a_2}^z=\pm 1,...,\sigma_{a_N}^z=\pm 1},\prod_{i=1}^{N}\sigma_{a_{i}}^z=1\right\},
\ee
where $\ket{\sigma^x=\pm 1}$ ($\ket{\sigma^z=\pm 1}$) are the $\pm 1$ eigenvectors of $\sigma^x$ ($\sigma^z$ respectively). Therefore, the model with $2N$ gapped boundary junctions has GSD $=2^{N-1}$. 

The two observations above have a physical interpretation via boundary anyon condensation. The smooth boundary segments prevent the rough boundary segments from exchanging condensate charges. When a bulk charge hops to a rough boundary segment, it condenses right there and cannot move along the boundary to other boundary segments. As a result, the condensate charge on the $i$-th rough boundary segment is a conserved quantity and is measured by $\sigma_{a_i}^z$. The global charge constraint further demands the total condensate charge to be trivial. 

\begin{figure}[h!]
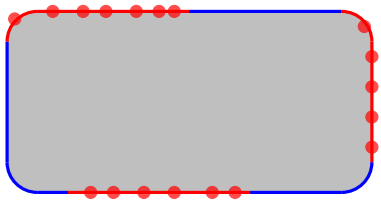

\centering
\IslandA
\caption{$6$ boundary segments with $3$ rough boundary segments (red regions) and $3$ smooth boundary segments (blue regions). Bulk charges can condense on the rough boundary segments but not on smooth boundary segments.}
\label{fig:island}
\end{figure}

We can visualize these ground-state quantum numbers: The bulk is like an ocean of charges, while the boundary segments are like the isolated islands in the ocean. Some bulk charges can land on these islands. Once the bulk charges land, they become condensate charges and can not travel along the boundary to other islands (see Fig. \ref{fig:island}). 

Therefore, the $2^{N-1}$-fold degenerate ground states are characterized\footnote{There are two sets of observables that commute with the Hamiltonian: (1) $\{\sigma_{a_i}^z\}$ representing the condensate charge on the $i$-th rough boundary segment, and (2) $\{\sigma_{a_i}^x\sigma_{a_j}^x\}$ representing a ribbon operator that hops through the bulk the condensate charge on the $i$-th rough boudnary segment to that on the $j$-th one. These two sets of observables do not commute with each other. We can choose either set to label the ground-state basis. We choose the first set in our paper.} by the configurations of independent condensate charges on the $N$ rough boundary segments subject to the global charge constraint $\prod_{i=1}^{N}\sigma_{a_{i}}^z=1$.  

To be precise, let us consider an example with $6$ gapped boundary junctions on a disk as shown in Fig. \ref{fig:shrinkgraph}. We use a $\Z_2$-valued variable $\sigma_{a_i}^z=\pm 1$ to represent the even or odd number of condensate charges on the rough boundary $i$. It is a topological invariant because it can not be changed by local operators on boundary $i$. The triples $(\sigma_{a_1}^z,\sigma_{a_2}^z,\sigma_{a_3}^z)$ subject to $\sigma_{a_1}^z\sigma_{a_2}^z\sigma_{a_3}^z=1$ describe the situations of charge condensations on three rough boundaries. The allowed configurations are $(1,1,1)$, $(-1,-1,1)$, $(-1,1,-1)$, and $(1,-1,-1)$. Thus the GSD $=4$. Generalizing to the case with $2N$ boundary junctions is straightforward. The $N$-tuples $(\sigma_{a_1}^z,\sigma_{a_2}^z,...,\sigma_{a_{N}}^z)$ subject to $\prod_{i=1}^Nz_i=1$ characterize the ground states. 

\section{The extended Levin-Wen model with gapped boundary junctions}\label{sec:ham}
Having seen the ELW $\Z_2$ model with gapped boundary junctions, we are now ready to construct a general exactly solvable model of topological orders with gapped boundary junctions based on the ELW model. In the end of Section \ref{sec:bjh}, one will see that the ELW $\Z_2$ model can be recovered as a special case of the general construction. Let us first briefly review the ELW model. We adopt the notations in \cite{Hu2017a}.

\subsection{Review of the ELW model}\label{sec:revw}
\begin{figure}[h!]
\centering
\includegraphics[scale=0.5]{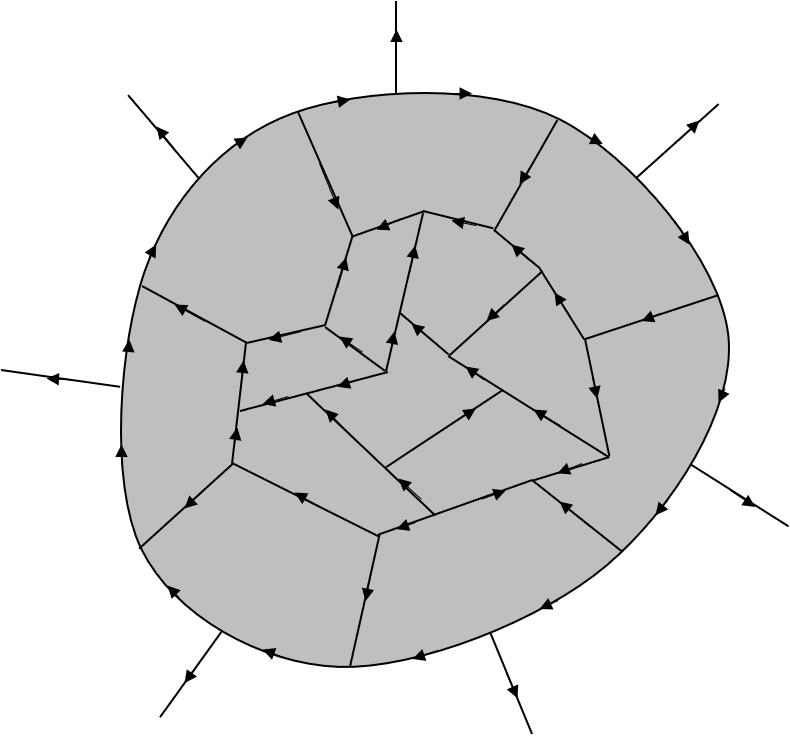}
\caption{A trivalent lattice $\Gamma$ of a disk with bulk in grey and boundary consisting of the dangling edges.}
\label{fig:lwbdGraph}
\end{figure}

\begin{figure}[h!]
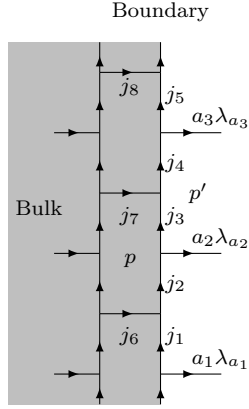

\centering
\bdryillu{a_1\lambda_{a_1}}{j_1}{j_2}{a_2\lambda_{a_2}}{j_3}{j_4}{a_3\lambda_{a_3}}{j_5}
\caption{A portion of a lattice with a boundary, which consists of the dangling edges labeled by $a_{i}\lambda_{a_i}$. The other edges are bulk edges. A bulk plaquette $p$ and a boundary open plaquette $p'$ are indicated.}
\label{fig:lwtails}
\end{figure}

The ELW model is a Hamiltonian lattice model defined on an oriented trivalent lattice $\Gamma$ with boundaries. An example lattice is depicted in Fig. \ref{fig:lwbdGraph}. Each edge is oriented. There are bulk edges and boundary edges---the dangling edges. There are also bulk plaquettes in grey and those boundary open plaquettes (see Fig. \ref{fig:lwtails}).

The input data of the model is a unitary fusion category (UFC) $\F$ and a Frobenius algebra $A$ in $\F$ that characterizes the gapped boundary condition for each boundary. A UFC $\F$ has a tensor description defined by a quadruple $\{L,N,d,G\}$. The set  $L=\{0,1,...,S\}$ collects the isomorphism classes of simple objects in $\F$. Each edge of the lattice carries an element of $L$. One also calls the elements of $L$ the string types. Each string type $j$ has a dual $j^*\in L$, which satisfies $j^{**}=j$. If we reverse the orientation of an edge and dual the corresponding string type $j\mapsto j^*$ at the same time, the state associated with the edge remains unchanged. There is always a trivial (unit) element $0\in L$. Note that $0^*=0$.

The fusion rule $N: L \times L \times L \rightarrow \mathbb{N}$ satisfies
\be\label{eq:fusion1}
N_{0 i}^{j} =N_{i 0}^{j}=\delta_{i j}, 
\ee
\be\label{eq:fusion2}
N_{i j}^{0} =\delta_{i j^{*}}, 
\ee
\be\label{eq:fusion3}
\sum_{x \in L} N_{i j}^{x} N_{x k}^{l} =\sum_{y \in L} N_{i y}^{l} N_{k j}^{y},
\ee
for any $i,j,k,l \in L$. For simplicity, we restrict the fusion rules to be multiplicity-free in this paper, such that $N_{i j}^k$ can be expressed by the delta function $N_{i j}^{k^*}:=\delta_{i j k}$ satisfying: $\delta_{i j k}=\delta_{j k i}$ and $\delta_{i j k}=\delta_{k^* j^* i^*}$. 

Given the fusion rule, we can define the quantum dimension $d:L \rightarrow \mathbb{R}$ that satisfies $d_j=d_{j^*}$ and 
\be\label{eq:qdim}
\sum_{k\in L}\delta_{i j k^*}d_k=d_id_j.
\ee
It is easy to see $d_0=1$.

The last element $G$ in the quadruple is the symmetric $6j$-symbol, which is a map $G: L^6 \rightarrow \mathbb{C}$  satisfying
\be
\begin{aligned}
G_{k l n}^{i j m}=G_{n k^{*} l^{*}}^{m i j}=G_{i j n^{*}}^{k l m^{*}} &=\alpha_{m} \alpha_{n} \overline{G_{l^{*} k^{*} n}^{j^{*} i^{*} m^{*}}}, \\
\sum_{n} \mathrm{~d}_{n} G_{k p^{*} n}^{m l q} G_{m n s^{*}}^{j i p} G_{l k r^{*}}^{j s^{*} n} &=G_{q^{*} k r^{*}}^{j i p} G_{m l s^{*}}^{r i q^{*}}, \\
\sum_{n} \mathrm{~d}_{n} G_{k p^{*} n}^{m l q} G_{p k^{*} n}^{l^{*} m^{*} i^{*}} &=\frac{\delta_{i q}}{\mathrm{~d}_{i}} \delta_{m l q} \delta_{k^{*} i p},
\end{aligned}
\ee
where the $\bar G$ is the complex conjugate, and $\alpha_j:=sgn(d_j)$.

For example, the fusion category $\rep_{\Z_2}$ has two simple objects $0$ and $1$, with $0^*=0$, $1^*=1$, and $d_0=d_1=1$. The fusion rule reads $\delta_{000}=1$ and $\delta_{011}=1$. The $6j$-symbol $G$ takes the form 
\be\label{eq:z26j}
G_{k l n}^{i j m}=\delta_{i j m} \delta_{k l m^{*}} \delta_{j k n^{*}} \delta_{i n l}.
\ee
Another fusion category to be considered in this paper is $\rep _{S_3}$, which has three self-dual simple objects corresponding to the three irreducible representations of $S_3$. These simple objects are denoted by $L=\{0,1,2\}$, with $d_0=1$, $d_1=1$, and $d_2=2$. The fusion rule reads
\be\label{eq:s3fusion}
\delta_{0 0 0}=1, \delta_{0 1 1}=1, \delta_{0 2 2}=1, \delta_{1 2 2}=1, \delta_{2 2 2}=1,
\ee
and the nonzero values of $G$ are 
\be\label{eq:s3sixj}
\begin{aligned}
&G_{000}^{000}=1, G_{111}^{000}=1, G_{222}^{000}=\frac{1}{\sqrt{2}}, G_{011}^{011}=1, G_{222}^{011}=\frac{1}{\sqrt{2}} \\
&G_{022}^{022}=\frac{1}{2}, G_{122}^{022}=\frac{1}{2}, G_{222}^{022}=\frac{1}{2}, G_{122}^{122}=\frac{1}{2}, G_{222}^{122}=-\frac{1}{2}.
\end{aligned}
\ee

A Frobenius algebra $A$ in $\F$ is a pair $(L_A,f)$, where $L_A$ is a set $\{(a,\lambda_a)\}$, and $f$ is the multiplication of $A$. In a pair $\{(a,\lambda_a)\}$, $a\in L$ is a string type, and $\lambda_a\in \{1,2,\dots,|a|\}$, where $|a|$ is the multiplicity of $a$ defined as the number of different pairs $(a, \lambda_a)\in L_A$ with the same $a$. To ease computation, we may write $(a, \lambda_a)$ as $a\lambda_a$ or $a_{\lambda_a}$ for short. The multiplication $f$ is a map : $L_A\times L_A\times L_A\rightarrow \mathbb{C}$ that satisfies
the following associativity and non-degeneracy conditions:
\begin{subequations}\label{eq:faCondi}
\be\label{eq:faAssoc}
\sum_{c\lambda_c}f_{a\lambda_a b\lambda_b c^*\lambda_c} f_{c\lambda_c r\lambda_r s^*\lambda_s } G^{abc^*}_{rs^*t}  v_c  v_t = \sum_{\lambda_t} f_{a\lambda_a t\lambda_t s^*\lambda_s}f_{b\lambda_b r\lambda_r t^*\lambda_t},
\ee
\be\label{eq:faNonDegen}
f_{b\lambda_b b^*\lambda_b 0_1} \neq0, \forall b\lambda_b \in L_A.
\ee
\end{subequations}
The Frobenius algebra $A$ defined above is also an object in $\F$ and can be expressed as $\oplus_{(a,\lambda_a)\in L_A}a_{\lambda_a}$.

For simplicity, in this paper, we assume $|0|\equiv 1$ and omit the multiplicity labels for those elements with multiplicity $\leq 1$. We also normalize the non-degeneracy condition as $f_{b\lambda_b b^{*}\lambda_b 0}=1, \forall b\in L_A$. A Frobenius algebra has the following properties: \be\begin{aligned}\label{eq:faprop}
\text{unit condition}&&& f_{b\lambda_{b} c^*\lambda_c 0}=\delta_{b\lambda_b, c\lambda_c},
\\
\text{cyclic condition}&&& f_{a\lambda_a b\lambda_b c\lambda_c}=f_{c\lambda_c a\lambda_a b\lambda_b},
\\
\text{strong condition}&&& \sum_{a\lambda_a, b\lambda_b}f_{a\lambda_a b\lambda_b c\lambda_c}f_{c^*\lambda_c b^*\lambda_b a^*\lambda_a}v_av_b=d_{A}v_c,
\end{aligned}
\ee
where $d_{A}=\sum_{b \lambda_b \in L_A} d_b$.

In $\F= \rep_{\Z_2}$ for example, there are two inequivalent Frobenius algebras: $A_1=0$, which is trivial, and $A_2= (L_{A_2}=\{0, 1\},f_{011}=1)$.

In $\F=\rep_{S_3}$, there are four inequivalent Frobenius algebras, listed in Table \ref{tab:Z23sols}.
\begin{table}[!h]
\small
\centering
\begin{tabular}{|c|}
\hline
Frobenius algebra (modulo Morita equivalence) in $\text{Rep}_{S_3}$\\ $f_{a\lambda_a b\lambda_b c\lambda_c}=f_{c\lambda_c a\lambda_a b\lambda_b}$, $f_{a\lambda_a a^*\lambda_a 0}=1$\\
\hline
$A_1=0$ \\ $A_2=0\oplus 1$ \\
$A_3=0\oplus2$, $f_{222}=-2^{-\frac{1}{4}}$  \\
$A_4=0\oplus1\oplus2_1\oplus2_2$, $f_{1 1 0}=1$, $f_{2_{2}12_{1}}=-i$, $f_{2_22_11}=i$,\ $f_{2_12_12_1}=-2^{-\frac{1}{4}}$, $f_{2_12_22_2}=2^{-\frac{1}{4}}$  
\\
\hline 
\end{tabular}
\caption{Frobenius algebras in $\rep_{S_3}$. }
\label{tab:Z23sols}
\end{table}

For each boundary edge of a boundary characterized by Frobenius algebra $A$, we assign an element in $ L_A$. 

The total Hamiltonian of the ELW model with a gapped boundary characterized by a Frobenius algebra $A$ defined  is 
\be\label{eq:lwBdryHamiltonian}
H=H_{\rm bulk}+H_{\rm bdry}= H_{\F} + H_{A},
\ee
where $H_{\F}$ is the bulk Hamiltonian term, and $H_{A}$ is the boundary Hamiltonian. 

The bulk Hamiltonian takes the form
\be\label{eq:lwHam}
H_{\F}=-\sum_{v} Q_v-\sum_{p} B_p,
\ee
where $Q_v$ and $B_p$ are the vertex and plaquette operators, and the sums run over all vertices $v$ and plaquettes $p$. The operator $Q_v$ acts on the local states at $v$ as
\be\label{eq:lwAvop}
Q_v\bket{\lwAv{i}{j}{k}}=\delta_{ijk}\bket{\lwAv{i}{j}{k}}.
\ee
The operators $B_p$ are defined as follows.
\be\begin{aligned}\label{eq:lwBpop}
B_p=&\frac{1}{D}\sum_{s\in L} d_sB_p(s),
\\
B_p(s)\bket{\lwBp{j_1}{j_2}{j_3}{j_4}{j_5}{j_6}} = &\sum_{j'_1,j'_2,j'_3,j'_4,j'_5,j'_6\in L} \left(\prod_{k=1}^6  v_{j_k} v_{j'_k}\right) G^{i_1j_1^*j_6}_{sj'_6j_1'^*} G^{i_2j_2^*j_1}_{sj_1'j_2'^*} G^{i_3j_3^*j_2}_{sj_2'j_{3}'^*}
\\ 
&\times  G^{i_4j_4^*j_3}_{sj'_3j_4'^*}G^{i_5j_5^*j_4}_{sj_4'j_5'^*}G^{i_6j_6^*j_5}_{sj_5'j_6'^*}\bket{\lwBp{j_1'}{j_2'}{j_3'}{j'_4}{j'_5}{j'_6}},
\end{aligned}\ee
where $v_i:=\sqrt{d_i}$. 

The boundary Hamiltonian is\footnote{There could exist another term describing the boundary charges as defined in Ref\cite{Hu2017a}. Nevertheless, in this paper we are not interested in the boundary charge excitations, and hence omit these terms.}
\be\label{eq:qdbdHamiltonian}
H_{A}=-\sum_{p'} \Bb_{p'},
\ee
where $p'$ runs over all boundary plaquettes. 
The $\overline B_{p'}$ is
\be\label{eq:lwbdBpop}
\Bb_{p'}=\frac{1}{d_A} \sum_{t\lambda_t \in L_A}  v_t \Bb_{p'}(t \lambda_t), 
\ee
where $\Bb_{p'}(t \lambda_t)$ acts on the open plaquette $p'$  as:
\begin{align}
\Bb_{p'}(t \lambda_t) \bket{\lwbdBp{j_{1}}{a_1 \lambda_{a_1}}{j_{2}}{j_3}{j_{4}}{a_{2} \lambda_{a_2}}{j_{5}}}&=\sum_{\substack{a_{2}'\lambda_{a_{2}'},a_1'\lambda_{a_1'}\in L_A\\ j_{4}',j'_{2}\in L}} f_{t^{*}\lambda_{t} a_{2}'\lambda_{a_2'} a_{2}^*\lambda_{a_2}}f_{a_1^{*}\lambda_{a_1} a'_1\lambda_{a_1'} t\lambda_{t}} u_{a_{1}}  u_{a_{2}} \times 
\nonumber \\
& u_{a_1'}u_{a_{2}'}G^{j_{5}^* j_{4} a_{2}}_{t^*a_{2}'j_{4}'}G^{j_{3} j_{2} j_{4}^*}_{t^*  j_{4}'^* j'_2}G^{j_{1}a_1j_{2}^* }_{t^*  j_2'^* a_1'} v_{j_2} v_{j_{4}}  v_{j_2'}  v_{j_{4}'} \bket{\lwbdBp{j_{1}}{a_1'\lambda_{a_1'}}{j'_{2}}{j_{3}}{j_{4}'}{a_{2}'\lambda_{a_2'}}{j_{5}}}. \label{eq:lwbdBptop}
\end{align}

All the operators $Q_v$, $B_p$, and $\Bb_{p'}$ in the  Hamiltonian are mutually commuting projectors, rendering the Hamiltonian \eqref{eq:lwBdryHamiltonian} exactly solvable. 
  
\subsection{Gapped boundary junction Hamiltonian}\label{sec:bjh}
In this section, we shall construct our model with gapped boundary junctions based on the ELW model. 

\begin{figure}[!h]
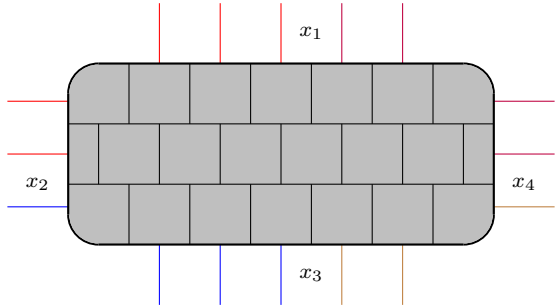

\centering
\Juc
\caption{A lattice (grey in the bulk) with $4$ gapped boundary junctions labeled by $x_1$, $x_2$, $x_3$, and $x_4$. The four different gapped boundary segments are in different colors. Edge orientation is omitted for simplicity.}
\label{fig:disklatt}
\end{figure}

Our model is defined on the same lattice as that of the ELW model; however, the boundary can have multiple segments, characterized by different Frobenius algebras in $\F$. A gapped boundary junction is an open boundary plaquette between two adjacent boundary segments (see Fig.\ref{fig:disklatt}). 
\begin{figure}[!h]
\centering
\jucbasis{j_1}{a_2\lambda_{a_2}}{j_2}{a_1\lambda_{a_1}}{j_3}
\caption{The local basis of the local Hilbert space at a gapped boundary junction. The bulk edges (black) are labeled by $j_1$, $j_2$, and $j_3$. The boundary edges (red and blue) are labeled by $a_1\lambda_{a_1}$ and $a_2\lambda_{a_2}$.}
\label{fig:jucab}
\end{figure}

The total Hamiltonian of our model now is that of the ELW model amended by a boundary junction Hamiltonian term. Namely,
\be\label{eq:totHam}
H_{\rm{tot}}=H_{\rm{bulk}}+H_{\rm bdry}+H_{\rm{junc}}=H_{\F}+\sum_{\text{seg }i}H_{\text{seg }i}+\sum_{\text{junc }x}H_{\text{junc }{x}},
\ee
where the index $i$ runs over all the boundary segments, and $x$ runs over all junctions. For each junction $x$, suppose the adjacent boundary segments are characterized by Frobenius algebras $A_1$ and $A_2$, the boundary junction Hamiltonian $H_{\text{junc }{x}}$ is defined by a projection operator, denoted by $\check{B}_x$, that commutes with $H_{\rm{bulk}}$ and $H_{\rm bdry}$. We present two equivalent definitions of $\check{B}_x$.

\noindent\textit{Definition 1.}
The $H_{\text{junc }{x}}=-\check B_{x}$ is defined as
\be\begin{aligned}\label{eq:bpone}
\check{B}_{x}\bket{\bponeA{j_1}{a_2\tilde{\lambda}_{a_2}}{j_2}{a_1\lambda_{a_1}}{j_3}}=&\mathcal{T}\left(\sum_{a'_1\lambda_{a'_1}\in L_{A_1}}\sum_{a'_2\tilde{\lambda}_{a'_2}\in L_{A_2}}\frac{\sqrt{D}u_{a'_1}u_{a'_2}}{u_{a_1} u_{a_2}}\bket{\bponeB{j_1}{a_2\tilde{\lambda}_{a_2}}{a'_2\tilde{\lambda}_{a'_2}}{j_2}{a_1\lambda_{a_1}}{a'_1\lambda_{a'_1}}{j_3}
}\right)
\\
=&\sum_{\substack{t\lambda_t\in L_{A_1},t\tilde{\lambda}_t\in L_{A_2}\\a'_1\lambda_{a'_1}\in L_{A_1}\\a'_2\tilde{\lambda}_{a'_2}\in L_{A_2}\\j'_2\in L}} f^{A_1}_{a^{*}_1\lambda_{a_1} t^*\lambda_{t} a'_1\lambda_{a'_1}}f^{A_2}_{t\tilde{\lambda}_t a^{*}_2\tilde{\lambda}_{a_2} a'_2\tilde{\lambda}_{a'_2}}G^{j_3^* j_2 a_1}_{t^* a'_1j'_2}G^{j'_2 a'^{*}_2 j^*_1}_{a_2 j^*_2 t}
\\
&\times v_tv_{j_2}v_{j'_2}u_{a_1}u_{a_2}u_{a'_1}u_{a'_2}\eta^{t}_{\lambda_t\tilde{\lambda}_t}\bket{\bponeA{j_1}{a'_2\tilde{\lambda}_{a'_2}}{j'_2}{a'_1\lambda_{a'_1}}{j_3}}
.
\end{aligned}
\ee
Here, the operation $\T$ is defined in Appendix \ref{sec:pach}, and the morphism $\eta: A_1\to A_2$ is defined by a function $\eta^{a}_{\lambda_{a}\tilde \lambda_{a}} \in \mathbb{C}$, $a\lambda_{a} \in L_{A_1}$, $a\tilde{\lambda}_{a} \in L_{A_2}$, satisfying:
\be\label{eq:ABmorcondi}
\T\left(\sqrt{D}\bket{\ABmorIII{a_1\lambda_{a_1}}{a_2\tilde \lambda_{a_2}}}\right)=\bket{\ABmorI{a_1\lambda_{a_1}}{a_2\tilde{\lambda}_{a_2}}}:=\eta^{a_1}_{\lambda_{a_1}\tilde{\lambda}_{a_2}}\delta_{a_1,a_2}\bket{\ABmorII{a_1\lambda_{a_1}}{a_2\tilde{\lambda}_{a_2}}},
\ee
where the red (blue) lines represent the elements of Frobenius algebra $A_1$ ($A_2$), the red (blue) dot represents the multiplication $f^{A_1}$ ($f^{A_2}$), and $\eta^{a_1}_{\lambda_{a_1}\tilde{\lambda}_{a_2}} \in \mathbb{C}$.

\noindent\textit{Definition 2.} 
The $H_{\text{junc }{x}}=-\check B_{x}$ is also defined as
\begin{align*}\small
&\check{B}_{x}\bket{\bponeA{j_1}{a_2\tilde{\lambda}_{a_2}}{j_2}{a_1\lambda_{a_1}}{j_3}}\\
=&\mathcal{T}\left(\sum_{a'_1\lambda_{a'_1}\in L_{A_1}}\sum_{a'_2\tilde{\lambda}_{a'_2}\in L_{A_2}}\frac{\sqrt{D}u_{a'_1}u_{a'_2}}{u_{a_1} u_{a_2}}\bket{\bponeBp{j_1}{a_2\tilde{\lambda}_{a_2}}{a'_2\tilde{\lambda}_{a'_2}}{j_2}{a_1\lambda_{a_1}}{a'_1\lambda_{a'_1}}{j_3}
}\right)
\\
=&\T\left(\sum_{\substack{k\lambda_k\in L_{A_1},k\tilde{\lambda}_k\in L_{A_2}\\k\tilde{\tilde{\lambda}}_k\in L_{A_{12}}\\a'_1\lambda_{a'_1}\in L_{A_1},a'_2\tilde{\lambda}_{a'_2}\in L_{A_2}}}\frac{\sqrt{D}u_{a'_1}u_{a'_2}}{u_{a_1} u_{a_2}} f^{A_1}_{a^{*}_1\lambda_{a_1} k^*\lambda_{k} a'_1\lambda_{a'_1}}f^{A_2}_{k\tilde{\lambda}_k a^{*}_2\tilde{\lambda}_{a_2} a'_2\tilde{\lambda}_{a'_2}}\right. \\
&\x \left.[\gamma^{12}]^k_{\lambda_k\tilde{\tilde{\lambda}}_k} [\bar \beta^{12}]^k_{\tilde{\tilde{\lambda}}_k\tilde\lambda_k}\bket{\bponeBpp{j_1}{a_2\tilde{\lambda}_{a_2}}{a'_2\tilde{\lambda}_{a'_2}}{j_2}{a_1\lambda_{a_1}}{a'_1\lambda_{a'_1}}{j_3}}\right)
\\
=&\sum_{\substack{k\lambda_k\in L_{A_1},k\tilde{\lambda}_k\in L_{A_2}\\ k\tilde{\tilde{\lambda}}_k\in L_{A_{12}} \\a'_1\lambda_{a'_1}\in L_{A_1}\\a'_2\tilde{\lambda}_{a'_2}\in L_{A_2}\\j'_2\in L}} f^{A_1}_{a^{*}_1\lambda_{a_1} k^*\lambda_{k} a'_1\lambda_{a'_1}}f^{A_2}_{k\tilde{\lambda}_k a^{*}_2\tilde{\lambda}_{a_2} a'_2\tilde{\lambda}_{a'_2}}G^{j_3^* j_2 a_1}_{k^* a'_1j'_2}G^{j'_2 a'^{*}_2 j^*_1}_{a_2 j^*_2 k}
\\
&\times v_kv_{j_2}v_{j'_2}u_{a_1}u_{a_2}u_{a'_1}u_{a'_2}[\gamma^{12}]^k_{\lambda_k\tilde{\tilde{\lambda}}_k} [\bar \beta^{12}]^k_{\tilde{\tilde{\lambda}}_k\tilde \lambda_k}\bket{\bponeA{j_1}{a'_2\tilde{\lambda}_{a'_2}}{j'_2}{a'_1\lambda_{a'_1}}{j_3}},\numberthis\label{eq:bponeDefII}
\end{align*}
where $\triangledown=\gamma^{12}: A_1\to A_{12}$, $\blacktriangledown=\beta^{12}: A_2\to A_{12}$, $\vartriangle=\bar{\gamma}^{12}: A_{12} \to A_1$, and $\blacktriangle=\bar{\beta}^{12}: A_{12}\to A_2$ define a common Frobenius subaglebra $A_{12}$ of $A_1$ and $A_2$, expressed by the tensors $[\gamma^{12}]^k_{\lambda_k\tilde{\tilde{\lambda}}_k}$, $[\beta^{12}]^k_{\lambda_k\tilde{\tilde{\lambda}}_k}$, $[\bar{\gamma}^{12}]^k_{\tilde{\tilde{\lambda}}_k\lambda_k}$, and $[\bar{\beta}^{12}]^k_{\tilde{\tilde{\lambda}}_k\lambda_k}$, satisfying:
\begin{align*}
\T&\left(\sqrt{D}\bket{\XYcondE{a_{1}\lambda_{a_{1}}}{a_{2}\tilde\lambda_{a_{2}}}}\right)=\bket {\XYcondF{a_{1}\lambda_{a_{1}}}{a_{2}\tilde\lambda_{a_{2}}}},
\\
&\bket{\XYcondI{a_{12}\lambda_{a_{12}}}{a'_{12}\lambda_{a'_{12}}}}=\delta_{a_{12}\lambda_{a_{12}},a'_{12}\lambda_{a'_{12}}}\bket{\XYcondJ{a_{12}\lambda_{a_{12}}}}, \numberthis \label{eq:cfade}
\\
&\bket{\XYcondK{a_{12}\lambda_{a_{12}}}{a'_{12}\lambda_{a'_{12}}}}=\delta_{a_{12}\lambda_{a_{12}},a'_{12}\lambda_{a'_{12}}}\bket{\XYcondJ{a_{12}\lambda_{a_{12}}}}.
\end{align*}

In general, there may be more than two gapped boundary segments. For three sequential segments, say, $A_{i-1},A_i$, and $A_{i+1}$, $A_i$ turns out to be an $A_{(i-1)i}$-$A_{i(i+1)}$-bimodule, with the corresponding module action given by
\be
[\tilde \rho_{i}^{\rm R}]_{a_i\lambda_{a_i} a'_i\lambda_{a'_i}}^{k \tilde \lambda_k}=\sum_{k\lambda_k\in L_{A_i}}f^{A_i}_{k^*\lambda_k a'^*_i\lambda_{a'_i}a_i\lambda_{a_i}}[\gamma^{i(i+1)}]^{k}_{ \lambda_k\tilde \lambda_k},
\ee
and 
\be[\tilde \rho_{i}^{\rm L}]_{a_{i}\lambda_{a_{i}} a'_{i}\lambda_{a'_{i}}}^{k\tilde \lambda_k}=\sum_{k\lambda_k\in L_{A_{i}}}f^{A_{i}}_{k\lambda_k a_i\lambda_{a_i} a'^*_i\lambda_{a'_i}}[\bar{\beta}^{(i-1)i}]^{k}_{\tilde \lambda_k \lambda_k}.
\ee
The $\check{B}_{x_i}$ operator \eqref{eq:bponeDefII} takes a more compact form:
\be
\begin{aligned}\label{eq:rewrite}
\check B_{x_i} \bket{\Oristate{a_i\lambda_{a_i}}{a_{i+1}\tilde{\lambda}_{a_{i+1}}}{x_i}}
=\T\left(\sum_{\substack{a'_i\lambda_{a'_i}\in L_{A_i}\\a'_{i+1}\tilde{\lambda}_{a'_{i+1}}\in L_{A_{i+1}}}}\frac{\sqrt{D}u_{a'_i}u_{a'_{i+1}}}{u_{a_i}u_{a_{i+1}}}\bket{\Rewrite{a_i\lambda_{a_i}}{a_{i+1}\tilde{\lambda}_{a_{i+1}}}{a'_i\lambda_{a'_i}}{a'_{i+1}\tilde{\lambda}_{a'_{i+1}}}}\right)
\end{aligned}
\ee

The two definitions of $\check{B}_x$ are equivalent. Given the morphism $\eta: A_1 \to A_2$, we can construct a common Frobenius subalgebra $A_{12}$ (see Appendix \ref{sec:CFSA} for the detail). Conversely, a common Frobenius subalgebra $A_{12}$, described by the projections and injections $\gamma^{12}:A_1\to A_{12}$, $\bar{\gamma}^{12}:A_{12}\to A_1$, $\beta^{12}:A_2\to A_{12}$, and $\bar{\beta}^{12}:A_{12}\to A_2$, yields a morphism $\eta=\bar{\beta}^{12}\circ 
\gamma^{12}:A_1\to A_2$.

Therefore, the total Hamiltonian \eqref{eq:lwBdryHamiltonian} is fully characterized by either of the following two sets of input data:
\begin{enumerate}
\item $\F$, $\{A_i\}$ for the $N$ gapped boundary segments $i=1,\cdots,N$, and $\{\eta^{i}\}$ for the gapped boundary junctions $x_i$.
\item $\F$, $\{A_i\}$ for the $N$ gapped boundary segments $i=1,\cdots,N$, and the Frobenius subalgebra $A_i\supseteq A_{i(i+1)}\subseteq A_{i+1}$ for the gapped boundary junctions $x_i$. We set  $A_{N( N+1)}:=A_{N1}$.
\end{enumerate}

It turns out that the operators $\check B_{x}$ are mutually commuting projectors and also commute with all other terms in the Hamiltonian. Thus, the Hamiltonian is exactly solvable. Proof of this statement is found in Appendix \ref{sec:comu}. 

We would like to stress that either the defining property \eqref{eq:ABmorcondi} of the morphism $\eta$ or \eqref{eq:cfade} of the common Frobenius subalgebra $A_{12}$ is a sufficient condition for the solubility of the model. Therefore, one can identify either Eq. \eqref{eq:ABmorcondi} or \eqref{eq:cfade} as the gapped boundary junction condition. This answers the question raised in the introduction.

Given a fusion category $\F$, all morphisms between the Frobenius algebras in $\F$ and common Frobenius subalgebras can be obtained by solving Eq. \eqref{eq:ABmorcondi} and Eq. \eqref{eq:cfade}.  For $\F=\rep_{\Z_2}$, the morphism between Frobenius algebras $0$ and $0\oplus1$  is trivial: $\eta^0_{11}=1$. For $\F=\rep_{S_3}$, the morphisms and common Frobenius algebras are shown in Table  \ref{tab:checkSthree}. The two structures $\eta$ and $A_{12}$ on each row of Table \ref{tab:checkSthree} are equivalent.
\begin{table}[!h]
\small
\centering
\begin{tabular}{|c|c|c|c|}
                \hline
 $A_1$   &  $A_2$ &  $\eta: A_1 \to A_2$ & $A_{12}$             
                \\
                \hline
               $0$  & $0\oplus 1$  & $\eta^0_{11}=1$ & $0$, $[\gamma^{12}]^0_{11}=1, [\bar\beta^{12}]^0_{11}=1$
               \\ \hline $0$ & $0\oplus 2$ & $\eta^0_{11}=1$ & $0$, $[\gamma^{12}]^0_{11}=1, [\bar\beta^{12}]^0_{11}=1$
               \\ \hline $0$  & $0\oplus 1\oplus 2_1\oplus 2_2$ & $\eta^0_{11}=1$ & $0$, $[\gamma^{12}]^0_{11}=1, [\bar\beta^{12}]^0_{11}=1$ 
               \\ \hline $0 \oplus 1$ &$0 \oplus 2$ & $\eta^0_{11}=1$ & $0$, $[\gamma^{12}]^0_{11}=1, [\bar\beta^{12}]^0_{11}=1$ 
\\ \hline 
 \multirow{2}{*}{
$0 \oplus 1$
}
 &  
  \multirow{2}{*}{
    $0\oplus 1\oplus 2_1\oplus 2_2$
}
& 
$\eta^0_{11}=\frac{1}{2},\, \eta^1_{11}=\frac{1}{2}$;
&
$0 \oplus 1$, $[\gamma^{12}]^a_{11}=1$,  $[\bar\beta^{12}]^a_{bc}=\eta^a_{bc}$;\\ 
\cline{3-4}
 & &
$\eta^0_{11}=\frac{1}{2}$, $\eta^1_{11}=-\frac{1}{2}$
&
$0 \oplus 1$, $[\gamma^{12}]^a_{11}=1$, $[\bar\beta^{12}]^a_{bc}=\eta^a_{bc}$
\\
\hline
\multirow{3}{*}{$0 \oplus 2$} & \multirow{3}{*}{$0\oplus 1\oplus 2_1\oplus 2_2$}  & 
$\eta^0_{11}=\frac{1}{3}, \, \eta^{2}_{11}=\frac{1}{3}, \, \eta^{2}_{12}=0$;
&
$0 \oplus 2$, $[\gamma^{12}]^a_{11}=1$, $[\bar\beta^{12}]^a_{bc}=\eta^a_{bc}$;
\\
\cline{3-4}
& &
$\eta^0_{11}=\frac{1}{3}, \, \eta^{2}_{11}=-\frac{1}{6}, \, \eta^{2}_{12}=-\frac{1}{2\sqrt{3}}$;
&
$0 \oplus 2$, $[\gamma^{12}]^a_{11}=1$, $[\bar\beta^{12}]^a_{bc}=\eta^a_{bc}$;
\\
\cline{3-4}
 & &
$\eta^0_{11}=\frac{1}{3}, \, \eta^{2}_{11}=-\frac{1}{6}, \, \eta^{2}_{12}=\frac{1}{2\sqrt{3}}$
&
$0 \oplus 2$, $[\gamma^{12}]^a_{11}=1$, $[\bar\beta^{12}]^a_{bc}=\eta^a_{bc}$
\\
\hline 
\end{tabular}
\caption{The morphisms and corresponding common Frobenius subalgebra $A_{12}$ between any two Frobenius algerbas $A_1$ and $A_2$ in $\rep_{S_3}$.}
\label{tab:checkSthree}
\end{table}

Equation \eqref{eq:ABmorcondi} may have multiple solutions. Two solutions $\eta$ and $\eta'$ are orthonormal if 
\be\label{eq:morthn}
\sum_{a_1\lambda_{a_1} \in L_{A_1}, a_2\tilde{\lambda}_{a_2} \in L_{A_2}}f^{A_1}_{a^*_1\lambda_{a_1} b\lambda_b a^*_{2}\lambda_{a_2}}f^{A_2}_{a_1\tilde{\lambda}_{a_1} a_2\tilde{\lambda}_{a_2} b^*\tilde{\lambda}_b}\eta^{a_1}_{\lambda_{a_1}\tilde{\lambda}_{a_1}}{\eta'}^{a_2}_{\lambda_{a_2}\tilde{\lambda}_{a_2}}\frac{v_{a_1} v_{a_2}}{v_b}=0
\ee 
for $b\lambda_b\in L_{A_1},b\tilde{\lambda}_b\in L_{A_2}$. A solution $\eta$ is minimal if there exists no two other nonzero orthonormal solutions $\eta_1$ and $\eta_2$, such that $\eta= \eta_1+\eta_2$. For the junction Hamiltonian to be well defined, we need to choose one particular minimal $\eta^{(x)}$ to construct $\check B_x$. With the gapped boundary conditions for all boundary segments fixed, there could be multiple gapped junction conditions. This further addresses the question raised in the introduction.

Let $\F=\rep_{\Z_2}$ as given by Eq. \eqref{eq:z26j}, the construction in this section reproduces the results in Sec \ref{sec:torc}, where all boundary segments are characterized by either of the two Frobenius algebras $L_{A_1}=\{0,1\}$ (rough) and $L_{A_2}=\{0\}$ (smooth), and all junctions are characterized by the trivial morphism between $A_1$ and $A_2$ given by $\eta^0_{11}=1$.

\section{ Ground states on a disk with gapped boundary junctions}\label{sec:top}
In this section, we study the ground states of our model on a disk with boundary junctions. We derive a GSD formula in terms of input data. We also write down the ground-state basis. 

\subsection{The GSD formula}
Our Hamiltonian model enables us to compute the GSD on a disk with $N$ gapped boundary junctions. These gapped boundary junctions divide the boundary of the disk into $N$ gapped boundary segments characterized by Frobenius algebras $A_1$, $A_2$,..., and $A_N$. 

For ground states, we can use a sequence of Pachner moves to remove all the bulk vertices and some boundary vertices of the lattice such that only one dangling edge for each gapped boundary segment remains (see either graph in Eq. \eqref{eq:GSD}).

On the simplified graph, the ground-state projector is
\be\label{eq:gpjunc}
P^0_{\rm junc}=B_p\prod_x \check B_x\prod_v Q_v,
\ee
where $p$ is the remaining bulk plaquette, $x$ runs over all the gapped boundary junctions, and $v$ runs over all the boundary vertices. 

The GSD is equal to the trace of $P^0_{\rm junc}$: 
\be
\label{eq:GSD}
\text{GSD}=\sum_{i=1}^{N}\sum_{\substack{a_i\lambda_{a_i} \in L_{A_i} \\ j_i\in L}}\Biggl\langle \evGSDdual \Biggr\vert P^0_{\rm junc}\Biggl\vert \evGSDdual \Biggr\rangle,
\ee
which evaluates to
\be\begin{aligned}\label{eq:evGSD}
\text{GSD}=&\sum_{i=1}^{N}\sum_{\substack{a_i\lambda_{a_i},a'_i\lambda_{a'_i}\in L_{A_i}\\ j_i,j'_i,s \in L\\t_{i(i+1)}\lambda_{t_{i(i+1)}}\in L_{A_i}\\ t_{i(i+1)}\tilde \lambda_{t_{i(i+1)}} \in L_{A_{i+1}}}}\frac{d_s}{D}v_{j_1}v_{j_2}\cdots v_{j_n}v_{j'_1}v_{j'_2}\cdots v_{j'_n}(T^{A_{1},A_{2},\eta^{(x_1)}})^{a_1\lambda_{a_1},a'_1\lambda_{a'_1},a_2\lambda_{a_2},a'_2\lambda_{a'_2}}_{j_n,j_1,j'_1,j_2,t_{12}\lambda_{t_{12}},t_{12}\tilde \lambda_{t_{12}}}
\\
&(T^{A_{2},A_{3},\eta^{(x_2)}})^{a'_2\lambda_{a'_2},a_2\lambda_{a_2},a_3\lambda_{a_3},a'_3\lambda_{a'_3}}_{j_1,j_2,j'_2,j_3,t_{23}\lambda_{t_{23}},t_{23}\tilde \lambda_{t_{23}}}  \cdots (T^{A_{n},A_{1},\eta^{(x_n)}})^{a'_n\lambda_{a'_n},a_n\lambda_{a_n},a'_1\lambda_{a'_1},a_1\lambda_{a_1}}_{j'_{n-1},j_n,j'_n,j'_1,t_{n1}\lambda_{t_{n1}},t_{n1}\tilde \lambda_{t_{n1}}}
\\
&G^{a^*_1 j'^*_n j'_1}_{s j_1 j^*_n}G^{a'^*_1 j'^*_1 j'_2}_{s j_2 j^*_1}G^{a^*_2 j'^*_2 j'_3}_{s j_3 j^*_2}\cdots G^{a'^*_n j'^*_{n-1} j'_n}_{s j_n j^*_{n-1}}, 
\end{aligned}\ee 
where we identify $i=N+1$ with $i=1$, and the tensor $T^{A,B,\eta}$ is defined by
\be\label{eq:factors}
(T^{A,B,\eta})^{a\lambda_{a},a'\lambda_{a'},b\lambda_{b},b'\lambda_{b'}}_{j_1,j_2,j'_2,j_3,t\lambda_t,t\tilde \lambda_t}:=f^A_{a\lambda_{a} t^*\lambda_{t} a'^*\lambda_{a'}}f^B_{t\tilde \lambda_t b\lambda_{b} b'^*\lambda_{b'}}G^{j_1^* j_2 a^*}_{t^* a'^* j'_2}G^{j'_2 b' j^*_3}_{b^* j^*_2 t}v_{j_2}v_{j'_2}v_{a}v_{b}v_t\eta^t_{\lambda_t\tilde \lambda_t}.
\ee
The GSD formula \eqref{eq:evGSD} takes a more compact form: 
\be\label{eq:gsdhomqe}
\text{GSD}=\text{dim}(\prod_x \check B_x {\rm Hom}(A_1\otimes A_2 \otimes A_3... \otimes A_N, 0)),  
\ee
where the hom-space is defined in the fusion category $\F$.

\subsection{Ground states}
In this subsection, we shall derive the basis of the ground-state Hilbert space and the corresponding quantum numbers of our model on a disk with $N$ gapped boundary junctions (see Fig. \ref{fig:gelwsim}(a)).

\subsubsection{Ground-state basis}\label{sec:gbasis}
\begin{figure}[h!]
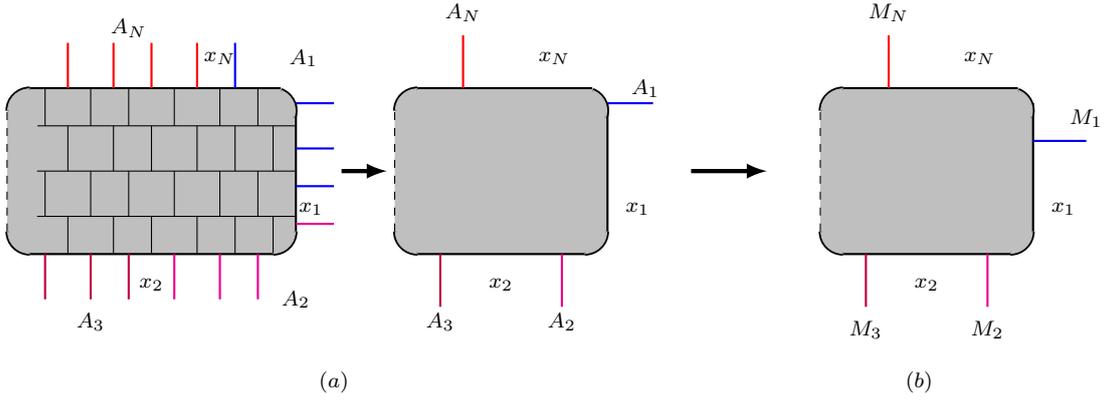

\centering
\gelwsim
\caption{(a) Simplification of the ELW model on a disk with $N$ boundary junctions. (b) Graph obtained by the operators $\check B_x$. Different colors represent distinct boundary segments.}
\label{fig:gelwsim}
\end{figure}

As previously done in Fig. \ref{fig:shrinkgraph}, we simplify the lattice in the left to that in the right of Fig. \ref{fig:gelwsim}(a) via Pachner moves, which preserve the ground-state Hilbert space. On this simplest reduced graph, each boundary segment has only one dangling edge. The $i$-th and the $(i+1)$-th dangling edges sandwich a gapped boundary junction labeled by $x_i$. We shall study the ground states on this reduced graph.

Let $M_i\subseteq A_i$ be an irreducible $A_{(i-1)i}$-$A_{i(i+1)}$-subbimodule of $A_i$. (As a $A_{(i-1)i}$-$A_{i(i+1)}$-bimodule, $A_i$ may be reducible. We say $M_i\subseteq A_i$ if $M_i$ is a direct summand appears in $A_i=M_i\oplus\cdots$.) For every tuple $(M_1\subseteq A_1,M_2\subseteq A_2,\cdots,M_N\subseteq A_N)$, there is a corresponding subspace $\mathcal{H}_{M_1,\cdots,M_N}$ of the ground-state Hilbert space, where the degree of freedom at each dangling edge $i$ is restricted to $m_i\lambda_{m_i}\in L_{M_i}$, such that $\mathcal{H}_{M_1,\cdots,M_N}$ is invariant under the $\check{B}_{x}$ operators for all $x$, see Fig. \ref{fig:gelwsim}(b). In other words, the ground-state Hilbert space is 
\be\label{eq:gdhs}
\mathcal{H}^{\text{GS}}=\bigoplus_{M_1,M_2,\cdots,M_N}\mathcal{H}_{M_1,\cdots,M_N}.
\ee
In what follows, we study $\mathcal{H}_{M_1,\cdots,M_N}$ and derive the ground-state basis. We shall act the gapped boundary junction operators $\check B_{x_1}$, $\check B_{x_2},\dots \check B_{x_N}$ on the junctions $x_1, x_2,\dots, x_N$ one by one in order.

Staring from boundary junction $x_1$ in Fig. \ref{fig:gelwsim}(b), by definition in Eq. \eqref{eq:bponeDefII}, $\check B_{x_1}$ acts as
\be\begin{aligned}\label{eq:projfusionBB}
&\check B_{x_1} \bket{\Oristate{m_1\lambda_{m_1}}{m_2\tilde{\lambda}_{m_2}}{x_1}}\\
=&\T\left(\sum_{\substack{m'_1\lambda_{m'_1}\in L_{M_1}\\m'_2\tilde{\lambda}_{m'_2}\in L_{M_2}}}\frac{\sqrt{D}u_{m'_1}u_{m'_2}}{u_{m_1}u_{m_2}}\bket{\projmorA{m_1\lambda_{m_1}}{m_2\tilde{\lambda}_{m_2}}{m'_1\lambda_{m'_1}}{m'_2\tilde{\lambda}_{m'_2}}}\right)
\\
=&\T\left(\sum_{k \in L} \sum_{\substack{m'_1\lambda_{m'_1}\in L_{M_1}\\m'_2\tilde{\lambda}_{m'_2}\in L_{M_2}}}\frac{\sqrt D v_k}{v_{m_1}v_{m_2}}M_{(m_1\lambda_{m_1}, m_2\tilde{\lambda}_{m_2})(m'_1\lambda_{m'_1}, m'_2\tilde{\lambda}_{m'_2})}^{M_1,M_2,A_{12},k}\bket{\projmorC{m'_1\lambda_{m'_1}}{m'_2\tilde{\lambda}_{m'_2}}{}{m_1}{m_2}}\right),
\end{aligned}
\ee
where for fixed $M_1$, $M_2$, $A_{12}$, and $k$, $M_{(m_1\lambda_{m_1}, m_2\tilde{\lambda}_{m_2})(m'_1\lambda_{m'_1}, m'_2\tilde{\lambda}_{m'_2})}^{M_1,M_2,A_{12},k}$ is a matrix:
\be\begin{aligned}\label{eq:multip}
M_{(m_1\lambda_{m_1}, m_2\tilde{\lambda}_{m_2})(m'_1\lambda_{m'_1}, m'_2\tilde{\lambda}_{m'_2})}^{M_1,M_2,A_{12},k}
=\sum_{\substack{t\lambda_t \in L_{M_1}\\t\tilde{\lambda}_t\in L_{M_2}\\t\tilde{\tilde{\lambda}}_t\in L_{A_{12}}}} 
&G_{m_2m'^*_2 k}^{m'^*_1 m_1 t^*}f^{A_1}_{m'^*_1\lambda_{m'_1}m_1\lambda_{m_1}t^*\lambda_t}f^{A_2}_{t\tilde{\lambda}_t m_2\tilde{\lambda}_{m_2} m'^*_2\tilde{\lambda}_{m'_2}}
\\
&
\times [\gamma^{12}]^t_{\lambda_t\tilde{\tilde{\lambda}}_t}[\bar{\beta}^{12}]^t_{\tilde{\tilde{\lambda}}_t\tilde{\lambda}_t}v_t u_{m'_1}u_{m'_2}u_{m_1}u_{m_2}.
\end{aligned}\ee

Because $\check B_{x_1}$ is a projector, the matrix $M^{M_1,M_2,A_{12},k}$ is also a projector. 

Denote the rank of $M^{M_1,M_2,A_{12},k}$ by $R_{M_1,M_2,A_{12},k}$.
Since the matrix $M^{M_1,M_2,A_{12},k}$ is a projector, there are a set $L_{M_{12}}=\{(k,\lambda_{k})|k\in L, \lambda_k \in \{1,2,\cdots,R_{M_1,M_2,A_{12},k}\}\}$ and a matrix $U_{\lambda_k(m_1\lambda_{m_1},m_2\lambda_{m_2})}^k$, such that
\be M_{(m_1\lambda_{m_1}, m_2\lambda_{m_2})(m'_1\lambda_{m'_1}, m'_2\lambda_{m'_2})}^{M_1,M_2,A_{12},k}=\sum_{\lambda_k}U_{\lambda_k (m_1\lambda_{m_1},m_2\lambda_{m_2})}^k[(U^k)^{\dag}]_{(m'_1\lambda_{m'_1},m'_2\lambda_{m'_2})\lambda_k},
\ee
and
\be\label{eq:Uidentity}
U^k (U^k)^\dag = \mathds{1}.
\ee

Define a right action tensor $\rho_{12}^{\rm R}$ by: 
\be\label{eq:rightAct}
\begin{aligned}(\rho_{12}^{\rm R})_{k\lambda_k k'\lambda_{k'}}^{t\tilde{\tilde{\lambda}}_t}=\sum_{\substack{m_1\lambda_{m_1}\in L_{M_1}\\m_2\lambda_{m_2},m'_2\lambda_{m'_2}\in L_{M_2}\\t\lambda_t\in L_{A_2}}}
&v_{m_2}v_{k'}U^k_{\lambda_{k'}(m_1\lambda_{m_1},m'_2\lambda_{m'_2})}[(U^k)^{\dag}]_{(m_1\lambda_{m_1},m_2\lambda_{m_2})\lambda_k}
\\
&\qquad
\x f_{m'^*_2\lambda_{m'_2}m_2\lambda_{m_2}t^*\lambda_t}^{A_2}G_{m^*_2m^*_1m'^*_2}^{k' tk^*}[\gamma^{23}]^t_{\lambda_t\tilde{\tilde{\lambda}}}.
\end{aligned}\ee
Similarly, we can define a left action tensor $\rho_{12}^{\rm L}$ by:
\be\label{eq:leftAct}
\begin{aligned}(\rho_{12}^{\rm L})_{k\lambda_k k'\lambda_{k'}}^{t\tilde{\tilde{\lambda}}_t}=\sum_{\substack{m_1\lambda_{m_1},m'_1\lambda_{m'_1}\in L_{M_1}\\m_2\lambda_{m_2}\in L_{M_2}\\t\lambda_t\in L_{A_{1}}}}
&v_{m_1}v_{k'}U^k_{\lambda_{k'}(m'_1\lambda_{m'_1},m_2\lambda_{m_2})}[(U^k)^{\dag}]_{(m_1\lambda_{m_1},m_2\lambda_{m_2})\lambda_k} 
\\
&\qquad
\x f_{t\lambda_t m_1\lambda_{m_1}m'^*_1\lambda_{m'_1}}^{A_1}G_{m_1m'^*_1m^*_2}^{k' k^*t^*}[\bar \beta^{N1}]^t_{\tilde{\tilde{\lambda}}\lambda_t}.
\end{aligned}\ee
Mathematically, $K_{12}:=(L_{M_{12}},\rho^{\rm{R}}_{12},\rho^{\rm{L}}_{12})$ is an $A_{N1}$-$A_{23}$-bimodule. The $\rho_{12}^{\rm R}$ and $\rho_{12}^{\rm L}$ will be used later.

Let $M_{12}\subseteq K_{12}$ be an irreducible $A_{N1}$-$A_{23}$-sub-bimodule of $K_{12}$. Via the $U$ matrix, we define unitary operators $T'^{M_{12}}_{2\rightarrow 1},T'^{M_{12}}_{1\rightarrow 2}$ by
\be\label{eq:projfusionI}
T'^{M_{12}}_{2\rightarrow 1}\bket{\FdecompositionA{m_1\lambda_{m_1}}{m_2\lambda_{m_2}}}=\T\left(\sum_{k\lambda_k\in L_{M_{12}}}\frac{\sqrt{D}v_{k}}{v_{m_1}v_{m_2}}U^k_{\lambda_k(m_1\lambda_{m_1},m_2\lambda_{m_2})}\bket{\FdecompositionD{m_1}{m_2}{k\lambda_k}}\right)
\ee
and
\be\label{eq:projfusionII}
T'^{M_{12}}_{1\rightarrow 2}\bket{\FdecompositionE{k\lambda_k}}=\T\left(\sum_{\substack{m_1\lambda_{m_1}\in L_{M_1}\\ m_2\lambda_{m_2}\in L_{M_2} }}[(U^k)^{\dag}]_{(m_1\lambda_{m_1},m_2\lambda_{m_2})\lambda_k}\bket{\FdecompositionC{k\lambda_k}{m_1\lambda_{m_1}}{m_2\lambda_{m_2}}}\right),
\ee
Then $\check B_{x_1}$ is decomposed as
\be
\check B_{x_1}=\sum_{M_{12}\subseteq K_{12}}T'^{M_{12}}_{1\to 2}\circ T'^{M_{12}}_{2\to 1},
\ee
where $T'^{M_{12}}_{1\to 2}\circ T'^{M_{12}}_{2\to 1}$ are orthonormal projections. Hence, the operators $T'^{M_{12}}_{2\rightarrow 1},T'^{M_{12}}_{1\rightarrow 2}$ for each $M_{12}$ preserve the ground-state Hilbert space.

\begin{figure}[h!]
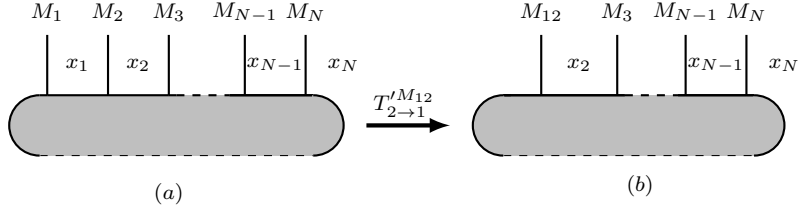

\centering
\groundbasisA
\caption{The action of $T'^{M_{12}}_{2\to 1}$ turns the reduced graph (a) to the graph (b).}
\label{fig:basisstepone}
\end{figure}

The action of ${T'}^{M_{12}}_{2\to 1}$ on $x_1$ turns the original reduced graph (Fig. \ref{fig:basisstepone}(a)) to the graph Fig. \ref{fig:basisstepone}(b). One can see that the original gapped junction $x_1$ is replaced by a dangling edge labeled by $M_{12}$.

The graph in Fig. \ref{fig:basisstepone}(b) is the starting point of the next step. We now focus on gapped boundary junction $x_2$. The operator $\check B_{x_2}$ induces a projection:
\be\label{eq:fusionFrothreeA}
\begin{aligned}
&\T\left(\sum_{\substack{k'\lambda_{k'} \in L_{M_{12}}\\m'_3\tilde{\lambda}_{m'_3}\in L_{M_3}}}\frac{\sqrt Du_{m'_3}u_{k'}}{u_{m_3}u_k}\bket{\FusionfroAA{k\lambda_k}{k'\lambda_{k'}}{m_3\tilde{\lambda}_{m_3}}{m'_3\tilde{\lambda}_{m'_3}}}\right)
\\
=&\T\left(\sum_{\substack{k'\lambda_{k'} \in L_{M_{12}}\\m'_3\lambda_{m'_3}\in L_{M_3}\\t\lambda_t\in L_{M_2}\\ t\tilde{\lambda}_t\in L_{M_3}\\t\tilde{\tilde{\lambda}}_t\in L_{A_{23}}}}\frac{\sqrt{D}u_{m'_3}u_{k'}}{u_{m_3}u_k}
f_{t\tilde{\lambda}_{t}m_3\tilde{\lambda}_{m_3}m'^*_3\tilde{\lambda}_{m'_3}}^{A_3}(\rho_{12}^{\rm R})_{k\lambda_k k'\lambda_{k'}}^{t\tilde{\tilde{\lambda}}_t}[\bar\beta^{23}]^{t}_{\tilde{\tilde{\lambda}}_t \tilde \lambda_t }\bket{\FusionfroBB{k\lambda_k}{k'\lambda_{k'}}{m_3\tilde{\lambda}_{m_3}}{m'_3\tilde{\lambda}_{m'_3}}{t\tilde \lambda_t}{t\tilde{\tilde{\lambda}}_t}}\right),
\end{aligned}
\ee

\begin{figure}[h!]
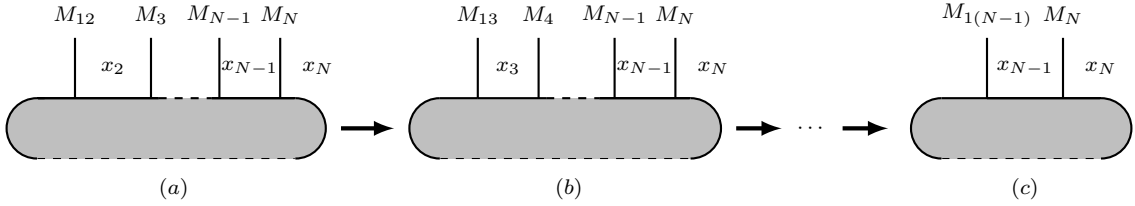

\centering
\groundbasisB
\caption{The intermediate steps.}
\label{fig:basisstepNM}
\end{figure}

We can then construct the operators similar to those in Eqs. \eqref{eq:projfusionBB} through \eqref{eq:projfusionII}, and obtain the graph in Fig. \ref{fig:basisstepNM}(b), in which the bimodule $M_{13}$ replaces the gapped junction $x_2$ in Fig. \ref{fig:basisstepNM}(a). We nevertheless do not detail these operators here. Let us continue similar operations on the remaining gapped junctions. After each operation, the original gapped junction $x_{i-1}$ is replaced by the bimodule $M_{1i}$. The subspaces $\mathcal{H}_{M_1,\cdots,M_N}$ are invariant under $\check B_{x_i}$. The operators $T'^{M_{1i}}_{2\to 1}$ and $T'^{M_{1i}}_{1\to 2}$ preserve the ground-state Hilbert space.
When we hit the last gapped boundary junctions $x_{N-1}$ and $x_N$, however, we need to consider both $\check B_{x_{N-1}}$ and $\check B_{x_N}$ at the same time, because the ground-states are simultaneous eigenvectors of $\check B_{x_{N-1}}=1$ and $\check B_{x_N}=1$. 

\begin{figure}[!h]
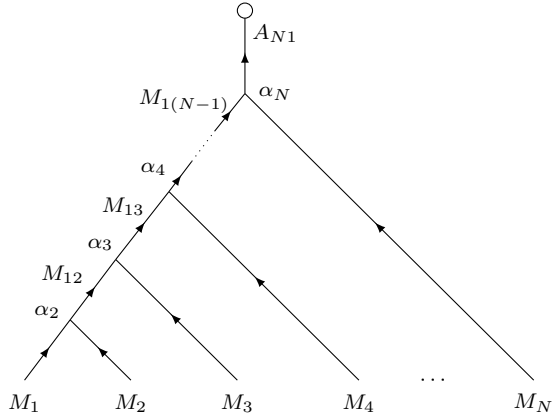

\centering
\fusiontreeAA
\caption{A fusion tree of the ground-state basis. The $A_{N1}$ is the trivial $A_{N1}$-$A_{N1}$-bimodule. The circle at the top is the counit $A_{N1}\to 0$. }
\label{fig:fusiontreeAA}
\end{figure}

In the operation at junction $x_i$ (for $i=2,\cdots,N-1$), we denote by $M_{1(i+1)}=M_{1i}\otimes_{A_{i(i+1)}}M_{i+1}$ the new $A_{N1}$-$A_{(i+1)(i+2)}$-bimodule, where $M_{1i}\otimes_{A_{i(i+1)}}M_{i+1}$ is a subobject of $M_{1i}\otimes M_{i+1}$ invariant under the projection $\check{B}_{x_{i}}:M_{1i}\otimes M_{i+1}\to M_{1i}\otimes M_{i+1}$. The operation $\otimes_{A_{i(i+1)}}$ is a bimodule morphism (see Appendix. \ref{sec:bimor} for details). The operators $T'^{M_{1i}}_{2\to 1}$ (for $i=2,\dots,N-1$) are topological observables of ground states. In general, there may be multiple $(T'^{M_{1i}}_{1\to 2},T'^{M_{1i}}_{2\to 1})$ for a fixed $M_{1i}$, and we denote such multiplicity degrees of freedom\footnote{Mathematically, the multiplicity degree of freedom $\alpha_i$ is defined by a linear independent basis of the bimodule morphism from $M_{1i}$ to $M_{1(i+1)}$.} by $\alpha_i$. The set $\{M_i,i=1,\dots,N;M_{1j},j=2,\dots,N-1;\alpha_k,k=2,\cdots,N\}$ labels the ground-state basis, namely, 
\begin{equation}\label{eq:gdbasis}
\left\{\left|\Phi_{M_1,M_2,\cdots,M_N;M_{12},M_{13},\cdots,M_{1(N-1)}};\alpha_2,\cdots,\alpha_N\right\rangle\right\}.
\end{equation}
We illustrate the ground-state basis \eqref{eq:gdbasis} by the fusion tree in Fig. \ref{fig:fusiontreeAA}.

\subsubsection{Boundary charge condensation}\label{sec:grdA}
Having derived the ground-state basis, we now study the physics of the ground-state basis via boundary charge condensation, which will be explained soon.

For our model on a disk with $N$ boundary segments, the general gapped boundary condition is now characterized by $\{A_i, A_{i(i+1)}| i=1,...,N\}$ with $A_{N(N+1)}:=A_{N1}$. In the bulk, there are two types of elementary bulk excitations: a charge (flux) excitation at a vertex $v$ (plaquette $p$) if $Q_v=0$ ($B_p=0$). Likewise, on the boundary, there are also two types of elementary boundary excitations---boundary charge (flux) excitations. When a bulk charge hops to the boundary, it will become either a nontrivial boundary charge or a trivial charge, i.e., vacuum. In the latter case, we say such a bulk charge is a condensate charge and condenses on the boundary. Particularly, on the $i$-th boundary segment characterized by $A_i$, the set of all possible condensate charges is $L_{A_i}$. 

The ground states are characterized by the allowed condensate charges on all boundary components. We will explain this in terms of a given gapped boundary condition $\{A_i,A_{i(i+1)} | i= 1, \dots,N\}$ as follows.
\begin{figure}[h!]
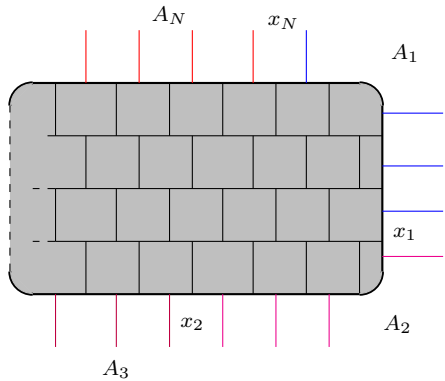

\centering
\gelw
\caption{An ELW model on a disk with $N$ boundary segments. Boundary edges of distinct boundary segments are painted in different colours.}
\label{fig:gelw}
\end{figure}

Consider a special case of our model without any junction, i.e., only one boundary segment. Suppose the boundary condition is characterized by the Frobenius algebra $A$. Then the allowed condensate charges on this boundary are $q\in L_{A}$. The total condensate charge on the boundary, however, is a conserved quantum number, and must be trivial. Hence, the ground state in this case is nondegenerate and is characterized by the trivial total condensate charge on the boundary. 

\begin{figure}[!h]
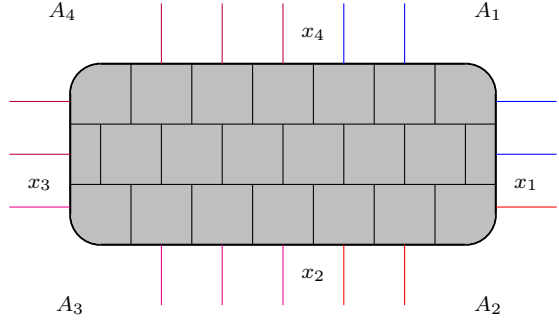

\centering
\fourjuncA
\caption{Our model with four boundary junctions. }
\label{fig:4juncgsd}
\end{figure}

If we allow multiple boundary segments separated by junctions, the ground states allow more possible condensate boundary charges. To be specific, suppose there are $N\in 2\Z$ boundary segments, as shown in Fig. \ref{fig:gelw}. Each boundary segment $i$ is characterized by a Frobenius algebra $A_i$, but $A_{2i}=0$ for $0\leq i\leq N/2$. For example, let $N=4$ and $A_1=A_3=A$ (see Fig.  \ref{fig:4juncgsd}.). The allowed condensate charges on the four boundary segments are labeled as
\begin{equation}\label{eq:fourcom}
q_1,q_3\in L_{A},
\quad
q_2=q_4=0.
\end{equation}
All these condensate charges must satisfy the global constraint that their total charge is trivial, which implies $\delta_{q_1q_30}=1$. The ground states are characterized by all possibilities of these condensate charges, rendering
\begin{equation}\label{eq:fourGSD}
\text{GSD}=\sum_{q_1,q_3\in L_A}\delta_{q_1q_30}.
\end{equation}
For arbitrary even $N$, the allowed condensate charges respectively on all boundary segments are 
\be\label{eq:caseA}
q_1\in L_{A_1}, \quad
q_3\in L_{A_3},...,q_{N-1}\in L_{A_{N-1}}, 
\quad
q_2=q_4=...=q_{N}=0.
\ee
These condensate charges must fuse to $0$, and the GSD is equal to $\dim \rm Hom(A_1\otimes A_3\otimes \cdots \otimes A_{N-1},0)$.

In the cases above, any boundary segment $i$ with a nontrivial Frobenius algebra is sandwiched between two boundary segments with trivial Frobenius algebra, which forbids nontrivial condensate charges. Therefore, the condensate charges on the boundary segment $i$ are prohibited to move to its adjacent segments. In general, however, condensate charges may move between adjacent boundary segments, as we now discuss. 

As mentioned earlier, for our model on a disk with $N$ boundary segments, the general boundary condition is characterized by $\{A_i,A_{i(i+1)} | i=1,\dots,N\}$. The subset $L_{A_i} \supseteq L_{A_{i(i+1)}} \subseteq L_{A_{i+1}}$ is the set of all condensate charges that are free to move between the $i$-th and the $(i+1)$-th boundary segment. We dub such condensate charges \textbf{mobile charges}. In contrast, the condensate charges in segment $i$ that are forbidden to move to the adjacent segments along the boundary are called \textbf{immobile charges}, which are labeled by the $M_i$ that appears in Eq. \eqref{eq:gdhs} (for an illustration see Fig. \ref{fig:islandC}).\footnote{Mathematically, bimodule categories are also called quotient categories.} The immobile charges are the topological quantum observables. Therefore, physically a ground state superposes the configurations of the immobile charges respectively in the $N$ boundary segments. This understanding leads to the following GSD formula.
\be
\text{GSD}=\sum_{M_1,\dots, M_N}\mathrm{dim}\,{\rm Hom}(M_1\otimes_{A_{12}}M_2 \otimes_{A_{23}} M_3 \dots \otimes_{A_{(N-1)N}}M_N, A_{N1}),
\ee
where the hom-space is defined in the category of $A_{N1}$-$A_{N1}$-bimodules, and  $M_{i}\otimes_{A_{i(i+1)}}M_{i+1}$ is a subobject of $M_{i}\otimes M_{i+1}$ invariant under the projection $\check{B}^{A_{i(i+1)}}_{x_{i}}:M_{i}\otimes M_{i+1}\to M_{i}\otimes M_{i+1}$. The dimension of the hom-space counts the multiplicity of $A_{N1}$ in the fusion of $M_1$, $M_2$, ..., and $M_N$. The $M_i$'s generalize the $q_i$'s in Eq. \eqref{eq:caseA}, in the sense that a $q_i$ is a $0$-$0$-bimodule.
\begin{figure}[h!]
\centering
\IslandC
\caption{An illustration of the mobile (black) and immobile (white and yellow) charges on a gapped boundary with four segments (red and blue). The black loop implies the mobility of the mobile charges.}
\label{fig:islandC}
\end{figure}
\section{Two examples}\label{sec:exams}

\noindent\textit{\textbf{Example 1}}. Consider $\F=\rm Rep_{\Z_2}$, and there are $6$ gapped boundary junctions on the lattice shown in Fig. \ref{fig:torsim} (a). The common Frobenius subalgebra of $A_1=0$ and $A_2=0\oplus 1$ is $A_{12}=0$. There are two irreducible $0$-$0$-bimodules: the trivial bimodule $0$ and the non-trivial bimodule $1$. The immobile charges are thus $0$ and $1$.

To write down the ground-state basis, we simplify the lattice to the one in Fig. \ref{fig:torsim} (b). As mentioned in Section \ref{sec:gbasis}, this simplification preserves the ground-state Hilbert space.  
\begin{figure}[h!]
\centering
\ToricSim
\caption{The original lattice (a) and the simplified lattice (b). Each blue (Red) boundary segment is characterized by the Frobenius algebra $A_1=0$ ($A_2=0\oplus 1$).}
\label{fig:torsim}
\end{figure}

According to Eq. \eqref{eq:gdbasis}, the ground-state basis in this example is illustrated by:  
\be\label{eq:fusiotreeC}
\fusiontreeC,
\ee
where the dashed lines are respectively labeled by $M_2$, $M_4$, $M_6$, and $A_{61}$, which can only be the $0$-$0$-bimodule $0$. The bimodules $M_1$, $M_3$, $M_5$, $M_{12}$, $M_{13}$, $M_{14}$, and $M_{15}$ can be either the trivial bimodule $0$ or non-trivial bimodule $1$. At each vertex, the fusion rules of $\F=\rep_{\Z_2}$ must be satisfied. The counit map is identity and thus omitted. We also omit the multiplicity labels $\alpha_i$ because they are identically unity.  

The fusion tree \eqref{eq:fusiotreeC} then results in exactly four allowed graphs:
\be\label{eqLfusionbasis}
\fusiontor,
\ee 
which indicate that there are only four ground-state basis vectors. Thus, $\text{GSD}=4$. 

\begin{figure}[!h]
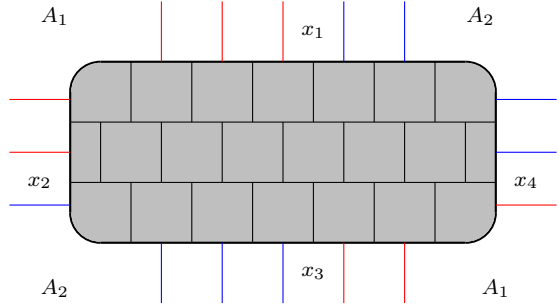

\centering
\fourjuncB
\caption{The lattice for Example 2. Each red (blue) boundary segment is characterized by the Frobenius algebra $A_1$ ($A_2$).}
\label{fig:4juncgsdB}
\end{figure}

\noindent\textit{\textbf{Example 2}}. Consider $\F=\rep_{S_3}$ and $4$ boundary junctions, as shown in Fig. \ref{fig:4juncgsdB}. There are six cases all told. We shall elaborate on two of them.  

Consider the case where $A_1 = 0\oplus 1$ and $A_2 =  0\oplus 1\oplus 2_1 \oplus 2_2$. The common Frobenius subalgebra of $A_1$ and $A_2$ is $A_{12}=0\oplus 1$ as shown in Table \ref{tab:checkSthree}. The set of mobile charges between any two adjacent segments is $L_{A_{12}}=\{0, 1\}$. There are three irreducible $A_{12}$-$A_{12}$-bimodules\footnote{Although $N_1$ and $N_2$ look like the same, they differ in module actions. The module actions are neglected because they serve no purpose here.}: $N_0 = 0\oplus 1$, $N_1=2$, and $N_2=2$.

Then, as $A_{12}$-$A_{12}$-bimodules, $A_1=N_0$, while $A_2= N_0\oplus N_1 \oplus N_2$, which identify the immobile charges in the corresponding boundary segments. The ground-state basis is illustrate by the following fuison tree.
\be
\fusionstree,
\ee
where $A_{41}=N_0$. The immobile charges $M_1$ and $M_3$ can only take $N_0$, whereas the immobile charges $M_2$ and $M_4$ and the internal degrees of freedom $M_{12}$ and $M_{13}$ take value in $\{N_0,N_1,N_2\}$. The multiplicity labels are omitted because they are identically one. There are three allowed basis vectors
\be
\label{eq:modefuseA}
\fusionreduB.
\ee
Therefore, GSD $=3$ in this case.

Now consider the case of $A_1=0\oplus 2$ and $A_2=0\oplus 1\oplus 2_1 \oplus 2_2$ in $\F=\rep_{S_3}$, the common Frobenius subalgebra is $A_{12}=0\oplus 2$ as shown in Table \ref{tab:checkSthree}.
The set of mobile charges between any two adjacent segments is $L_{A_{12}}=\{0, 2\}$. There are three irreducible $A_{12}$-$A_{12}$-bimodules: $N'_0 = 0\oplus 2$, $N'_1=1\oplus 2$, and $N'_2=2\oplus 2$. 

Then, as $A_{12}$-$A_{12}$-bimodules, $A_1=N'_0$, while $A_2= N'_0\oplus N'_1$, which identify the immobile charges in the corresponding boundary segments. The ground-state basis is illustrated by the following fuison tree.
\be\label{eq:efffuse1}
\fusionreduA,
\ee
where $A_{41}=N'_0$. The immobile charges $M_1$ and $M_3$ can only take $N'_0$, while the immobile charges $M_2$ and $M_4$ and the internal degrees of freedom $M_{12}$ and $M_{13}$ take value in $\{N'_0,N'_1\}$. The multiplicity labels are omitted because they are identically one. There are two allowed basis vectors 
\be\label{eq:modefuseB}
\fusionreduD.
\ee
Thus, GSD$=2$ in this case. 

Similarly but not to be elaborated, we can obtain the GSDs for the other choices of $A_1$ and $A_2$ in this example. Table \ref{tab:4juncgsd} records the results for all cases.
\begin{table}[!h]
\small
\centering
\begin{tabular}{|c|c|c|}
                \hline
Frobenius algebra $A_1$   & Frobenius algebra $A_2$ &  GSD             
                \\
                \hline
               $0$  & $0\oplus 1$  & $2$
               \\ \hline $0$ & $0\oplus 2$ & $2$
               \\ \hline $0$  & $0\oplus 1\oplus 2_1\oplus 2_2$ & $6$
               \\ \hline $0 \oplus 1$ &$0 \oplus 2$ & $6$
\\ \hline $0\oplus 1$ & $0\oplus 1\oplus 2_1\oplus 2_2$ & $3$
\\
\hline
$0 \oplus 2$ & $0\oplus 1\oplus 2_1\oplus 2_2$ & $2$
\\
\hline 
\end{tabular}
\caption{The GSDs in all cases of Example 2. }
\label{tab:4juncgsd}
\end{table}

\section{Discussion}\label{sec:con}
In the ground states of our model, there is a boundary defect at junction $x_i$ identified by an irreducible $A_i$-$A_{i+1}$-bimodule $X_{i(i+1)}:=(A_i \otimes A_{i+1}, \check B_{x_i})$, where $\check B_{x_i}$ is a projector
\be\begin{aligned}\label{eq:proexci}
& \check B_{x_i} \bket{\Oristate{a_i\lambda_{a_i}}{a_{i+1}\lambda_{a_{i+1}}}{x_i}}\\
=&\T\left(\sum_{\substack{a'_i\lambda_{a'_i}\in L_{A_i}\\a'_{i+1}\lambda_{a'_{i+1}}\in L_{A_{i+1}}}}\frac{\sqrt{D}u_{a'_i}u_{a'_{i+1}}}{u_{a_i}u_{a_{i+1}}}\bket{\projexci{a_i\lambda_{a_i}}{a_{i+1}\lambda_{a_{i+1}}}{a'_i\lambda_{a'_i}}{a'_{i+1}\lambda_{a'_{i+1}}}}\right).
\end{aligned}
\ee
Here, the tensor $\eta$ is defined in Eq. \eqref{eq:ABmorcondi}. The irreducible $A_i$-$A_{i+1}$-bimodule $X_{i(i+1)}$ is the sub-object of $A_i\otimes A_{i+1}$ invariant under $\check B_{x_i}$. 

In general, elementary junction excitations (including ground states) are local eigenvectors of $\check B_{x_i}$ in Eq. \eqref{eq:bpone}, and hence are labeled by irreducible $A_i$-$A_{i+1}$-bimodules. An excitation is trivial (i.e., ground state) if the corresponding bimodule is $X_{i(i+1)}$, and the ground-state Hilbert space can be expressed as
\begin{equation}\label{eq:HomMulti}
\mathrm{Hom}(X_{12}\otimes X_{23} \otimes \cdots \otimes X_{N1},A_1),
\end{equation}
where the Hom-space is defined in the multi-fusion category of bimodules over the Frobenius algebras. This classification of boundary junction excitations agrees with that of boundary defects in Ref. \cite{Cong2017a} because categories of bimodules are equivalent to module functor categories \cite{Ostrik2003}.

\begin{figure}[h!]
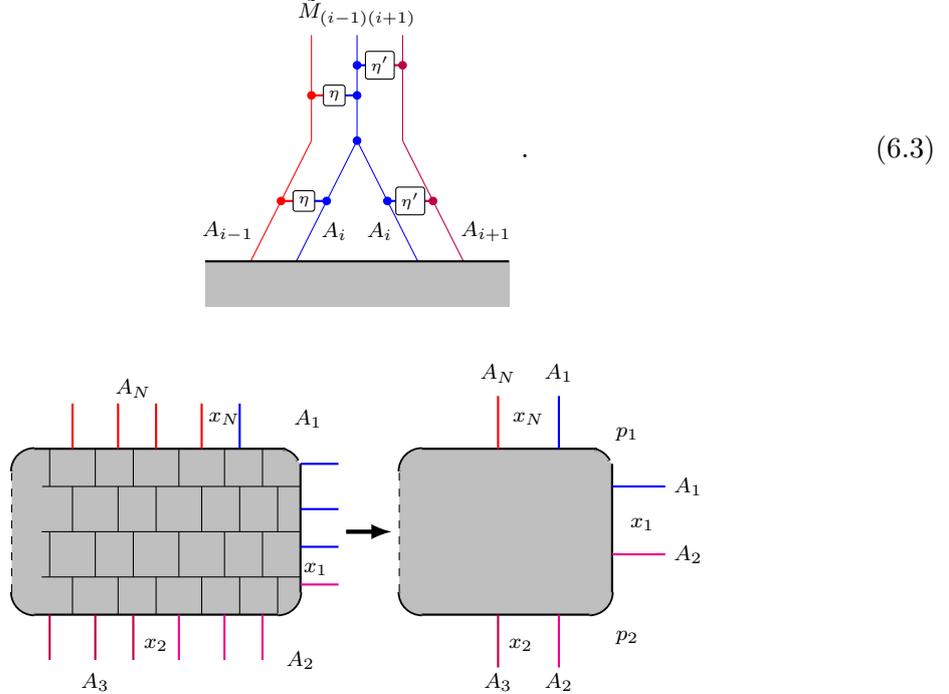

\centering
\exciA
\caption{Simplifications of the ELW model on a disk with $N$ boundary junctions. The boundary junctions and boundary plaquettes are labeled by  $x_i$ and $p_i$ respectively in the reduced graph.}
\label{fig:exciA}
\end{figure}

Two neighbouring boundary defects can fuse to a boundary defect. On the lattice, this fusion is realized by the procedure of reducing the number of open edges on a boundary segment to just two, as shown in Fig. \ref{fig:exciA}. The fusion process is described by the morphism $X_{(i-1)i} \otimes X_{i(i+1)} \to \tilde M_{(i-1)(i+1)}$, where $\tilde M_{(i-1)(i+1)}$ is an $A_{i-1}$-$A_{i+1}$-bimodule. This morphism is depicted as 
\be\label{eq:diamfusion}
\bimodulefusion.
\ee

\acknowledgments
YW is supported by NSFC grant No. 11875109, General Program of Science and Technology of Shanghai No. 21ZR1406700, Fudan University Original Project (Grant No. IDH1512092/009), and Shanghai Municipal Science and Technology Major Project (Grant No.2019SHZDZX01).

\appendix

\section{ Pachner moves and graphical tools}\label{sec:pach}
Any two trivalent graphs $\Gamma$  and $\Gamma'$ with the same topology can be mutated into each other by a composition of  elementary moves---the two-dimensional Pachner moves.

In the bulk, there are three Pachner moves, which induce the following unitary linear maps: 
\be\label{eq:pachI}
\begin{aligned}
&T_{2 \rightarrow 2} \bket{\pachA{j_1}{j_2}{j_3}{j_4}{j_5}}=\sum_{j'_5}G^{j_1 j_2 j_5}_{j_3 j_4 j_5'}v_{j_5}v_{j'_{5}} \bket{\pachB{j_1}{j_2}{j_3}{j_4}{j'_5}},
\\
&T_{1 \rightarrow 3} \bket{\pachC{j_1}{j_2}{j_3}}= \sum_{j_4,j_5,j_6}\frac{v_{j_4}v_{j_5}v_{j_6}}{\sqrt{D}} G^{j_{1} j_2 j_3}_{j_5 j_6 j_4}\bket{\pachD{j_1}{j_2}{j_3}{j_4}{j_5}{j_6}},
\\
&T_{3 \rightarrow 1} \bket{\pachD{j_1}{j_2}{j_3}{j_4}{j_5}{j_6}}=\frac{v_{j_4}v_{j_5}v_{j_6}}{\sqrt{D}}G^{j_1^*j_3^*j_2^*}_{j_5j_4^*j_6^*}\bket{\pachC{j_1}{j_2}{j_3}}.
\end{aligned}
\ee
The moves $T_{3 \rightarrow 1}$ and $T_{2 \rightarrow 2}$ can combine to define a move to eliminate bubbles: 
\be\label{eq:bubble}
T_{2 \rightarrow 0}\bket{\pachE{k}{i}{j}{k'}}=\delta_{k k'}\frac{v_i v_j}{v_k \sqrt{D}}\bket{\pachF{k}}.
\ee 
There are also two boundary Pachner moves, inducing the following two unitary linear maps: 
\be\label{eq:pachII}
\begin{aligned}
&T_{1 \rightarrow 2} \bket{\pachbA{j_1}{a_1\lambda_{a_1}}{j_2}}=\sum_{a_2\lambda_{a_2}, a_3\lambda_{a_3}}\frac{f_{a_1^*\lambda_{a_1} a_2\lambda_{a_2} a_3\lambda_{a_3}}u_{a_2}u_{a_3}}{u_{a_1}\sqrt{d_{A}}}\bket{\pachbB{j_1}{a_1\lambda_{a_1}}{j_2}{a_2\lambda_{a_2}}{a_3\lambda_{a_3}}},\\
&T_{2 \rightarrow 1}\bket{\pachbC{j_1}{a_1\lambda_{a_1}}{j_2}{a_2\lambda_{a_2}}{j_3}}=\sum_{a_3\lambda_{a_3}}\frac{f_{a^*_1\lambda_{a_1} a_3\lambda_{a_3} a_2^*\lambda_{a_2}}u_{a_3}\sqrt{D}}{u_{a_1}u_{a_2}\sqrt{d_A}}\bket{\pachbD{j_1}{a_1}{j_2}{a_2}{j_3}{a_3\lambda_{a_3}}}.
\end{aligned}
\ee
For convenience, we shall refer to all these linear maps as the $T$ maps. Under any composition of the $T$ maps, the ground-state Hilbert space is invariant. This fact relies on the algebraic properties of $6j$-symbols and Frobenius algebras multiplications $f$.

In general, to deform an intial graph state to a final one using the $T$ maps has more than one ways. To avoid this non-uniqueness, as a convention, we can label in the initial graph the plaquettes to be annihilated with a cross '$\times$', while in the final graph the plaquettes to be created with a dot '$\cdot$'. We can then specify the unique transformation, denoted by $\mathcal{T}$, between any two graph states, independent of the choice and ordering of the $T$ maps comprising $\mathcal{T}$. 

In terms of the $T$ maps, for example, $B_p$ can be written as
\be\begin{aligned}\label{eq:bbp}
B_p=&\mathcal{T}\left(\bbpB\right)
\\
=& T_{3\rightarrow1} T_{2\rightarrow2} T_{2\rightarrow2} T_{2\rightarrow2} T_{2\rightarrow2} T_{1\rightarrow3}.
\end{aligned}\ee

We adopt the thick-line convention in Ref. \cite{Hu2017a}. That is, a thick line indicates a summation over the input Frobenius algebra, and a thick dot represents the multiplication $f$ in the Frobenius algebra. We can then express the associativity and strong condition of a Frobenius algebra as:
\be\label{eq:grasso}
\mathcal{T}(\bket{\frassoA{a_1 \lambda_{a_1}}{a_2 \lambda_{a_2}}{a_3 \lambda_{a_3}}{a_4 \lambda_{a_4}} })=\bket{\frassoB{a_1 \lambda_{a_1}}{a_2 \lambda_{a_2}}{a_3 \lambda_{a_3}}{a_4 \lambda_{a_4}}},
\ee
\be\label{eq:grstrong}
\mathcal{T}(\frac{\sqrt{D}}{d_A}\bket{\frstrongA{c \lambda_c}{c \lambda_c}})=\bket{\frstrongB{c \lambda_c}}.
\ee
The $\Bb_p$ operator can be expressed as:
\be\label{eq:bpgr}
\Bb_p \bket{\lwbdBp{j_{1}}{a_1}{j_{2}}{j_3}{j_{4}}{a_{2}}{j_{5}}}=\mathcal{T}(\sum_{a'_1, a'_2}\frac{\sqrt{D}u_{a'_1}u_{a'_2}}{d_Au_{a_1}u_{a_2}}\bket{\lwbdgr{j_{1}}{a_1}{j_{2}}{j_3}{j_{4}}{a_{2}}{j_{5}}{a'_2}{a'_1}}).
\ee
\section{Some properties of $\check B_x$}\label{sec:comu}
In the main text, we state that $\check B_{x}$ are projectors and commute with the other Hamiltonian terms $Q_v$, $B_p$, and $\Bb_p$. Here, we sketch the proof. We consider the definition of $\check B_x$ in Eq. \eqref{eq:bpone}. We use red(blue) liens to represent elements in Frobenius algebra $A$($B$).

First, we first check $[\check{B}_{x}, B_{p}]=0$. The only nontrivial case is when $p$ is adjacent to $x$. The product of $B_{p}$ and $\check{B}_{x}$ can be written as a composition of the $T$ maps defined in Appendix \ref{sec:pach}. Namely
\be
\begin{aligned}\label{eq:bpcbp}
\check{B}_{x}B_{p}=&\sum_{a'\lambda_{a'}}\sum_{b'\lambda_{b'}}D^{\frac{1}{2}} \frac{u_{a'}u_{b'}}{u_au_b}\mathcal{T}\left(\bpcbpt\rightarrow\bpcbp\rightarrow\bpcbptt\right).
\\
=&B_{p}\check{B}_{x}.
\end{aligned}
\ee 
The second equality above holds because as mentioned in Appendix \ref{sec:pach}, in our convention, the ordering of the $T$ maps comprising a $\mathcal{T}$ is irrelevant.

Second, we check $[\check{B}_{x},\Bb_{p'}]=0$. This is obvious when the two plaquettes $x$ and $p'$ are not adjacent. Otherwise, we have
\begin{align*}
\Bb_{p'}\check{B}_{x}\bket{\bbpcbbpA}=&\mathcal{T}\left(\sum_{a'_1\lambda_{a'_1}}\sum_{a'_2\lambda_{a'_2}}\sum_{b'\lambda_{b'}}\frac{Du_{a'_1}u_{a'_2}u_{b'}}{d_Au_{a_1}u_{a_2}u_{b}}\bket{\bbpcbbpB}\right)
\\
=&\mathcal{T}\left(\sum_{a'_1\lambda_{a'_1}}\sum_{a'_2\lambda_{a'_2}}\sum_{b'\lambda_{b'}}\frac{Du_{a'_1}u_{a'_2}u_{b'}}{d_Au_{a_1}u_{a_2}u_{b}}\bket{\bbpcbbpC}\right) \numberthis \label{eq:bbpc}
\\
=&\check{B}_{x}\Bb_{p'}\bket{\bbpcbbpA},
\end{align*}
where the second equality follows Eq. \ref{eq:grasso}.  

Third, since a $Q_v$ is a diagonal matrix with entries being 1 and 0, it commutes with $\check B_x$. 

Finally, we verify that each $\check{B}_{x}$ is a projector. 
\begin{align*}
\check{B}_{x}\check{B}_{x}\bket{\jucproA}=&\check{B}_{x}\mathcal{T}\left(\sum_{a'\lambda_{a'}}\sum_{b'\lambda_{b'}}\frac{\sqrt{D}u_{a'}u_{b'}}{u_{a}u_{b}}\bket{\jucproB{a'\lambda_{a'}}{b'\lambda_{b'}}}\right)
\\
=&\mathcal{T}\left(\sum_{a''\lambda_{a''}}\sum_{b''\lambda_{b''}}\frac{Du_{a''}u_{b''}}{u_{a}u_{b}}\bket{\jucproC}\right)
\\
=&\mathcal{T}\left(\sum_{a''\lambda_{a''}}\sum_{b''\lambda_{b''}}\frac{Du_{a''}u_{b''}}{u_{a}u_{b}}\bket{\jucproD}\right)\numberthis \label{eq:jucpro}
\\
=&\mathcal{T}\left(\sum_{a''\lambda_{a''}}\sum_{b''\lambda_{b''}}\frac{\sqrt{D}u_{a''}u_{b''}}{u_{a}u_{b}}\bket{\jucproB{a''\lambda_{a''}}{b''\lambda_{b''}}}\right)\\
=&\check{B}_{x}\bket{\jucproA},
\end{align*}
where the third and fourth equalities are due to Eqs. \ref{eq:grasso} and \eqref{eq:ABmorcondi}.

\section{Common Frobenius subalgebra}\label{sec:CFSA}

Here, we show that the \textit{Definition 1} in Eq. \eqref{eq:bpone} and \noindent\textit{Definition 2} in Eq. \eqref{eq:bponeDefII} are equivalent.

Given a common Frobenius algebra $A_{12}$ of $A_1$ and $A_2$ defined by $\gamma^{12}$, $\bar \gamma^{12}$, $\beta^{12}$, and $\bar \beta^{12}$ in Eq. \eqref{eq:cfade}, if we define $\eta:=\bar\beta^{12}\circ\gamma^{12}$, the consistency conditions in Eq. \eqref{eq:cfade} imply the defining condition of $\eta$ in Eq. \eqref{eq:ABmorcondi}. Therefore, each $A_{12}$ results in an $\eta$.

Conversely, for each morphism $\eta : A_1 \to A_2$, there exists $A_{12}$, $\triangledown=\gamma^{12}: A_1\to A_{12}$, $\blacktriangledown=\beta^{12}: A_2\to A_{12}$, $\vartriangle=\bar{\gamma}^{12}: A_{12} \to A_1$, and $\blacktriangle=\bar{\beta}^{12}: A_{12}\to A_2$, such that
\be
\label{eq:compactSVD}
\bket{\singdA{a_1\lambda_{a_1}}{a_2\lambda_{a_2}}}=\bket{\singdB{a_1\lambda_{a_1}}{a_2\lambda_{a_2}}},
\ee
where 
\begin{align*}
\beta^{12}\bar\beta^{12}=\mathbbm{1}, \quad \gamma^{12}\bar\gamma^{12}=\mathbbm{1},
\end{align*}
and $\zeta:A_{12}\to A_{12}$ is an isomorphism. Equation \eqref{eq:compactSVD} is a compact singular value decomposition.  

Let $\tilde{\beta}^{12}=\zeta^{-1}\circ\beta^{12}$ and $\bar{\tilde{\beta}}^{12}=\bar{\beta}^{12}\circ\zeta$. The maps $\gamma^{12}$, $\bar \gamma^{12}$, $\tilde{\beta}^{12}$, and $\bar{\tilde{\beta}}^{12}$ satisfy all the conditions in Eq. \eqref{eq:cfade}. Therefore, we arrive at \textit{Definition 2}.

\section{Bimodule morphisms} \label{sec:bimor}
Let $A$ and $B$ be two inequivalent Frobenius algebras in a fusion category $\F$. An $A$-$B$ bimodule is a set $L_M:=\{(m,\lambda_m)|m\in L, \lambda_m \in \mathbb{N}\}$, equipped with an action $P_{m\lambda_m j\lambda_j n\lambda_n}^{a\lambda_a b\lambda_b}$ satisfying
\be\label{eq:bimod}
\mathcal{T}(\bket{\bimoduleA{m\lambda_m}{a_1\lambda_{a_1}}{b_1\lambda_{b_1}}{a_2\lambda_{a_2}}{b_2\lambda_{b_2}}{n\lambda_n}})=\bket{\bimoduleB{m\lambda_m}{a_1\lambda_{a_1}}{b_1\lambda_{b_1}}{a_2\lambda_{a_2}}{b_2\lambda_{b_2}}{n\lambda_n}},
\ee
where the boxed $P_M$ means
\be\label{eq:bimodf}
\bket{\bimoduleC{m\lambda_m}{a\lambda_{a}}{b\lambda_b}{n\lambda_n}}\equiv\sum_{j\lambda_j\in L_M}P_{m\lambda_m j\lambda_j n\lambda_n}^{a\lambda_a b\lambda_b}\bket{\bimoduleD{m\lambda_m}{a\lambda_a}{b\lambda_b}{n\lambda_n}}.
\ee
  
By definition \eqref{eq:bimod}, $A \otimes B$ is automatically an $A$-$B$-bimodule with the action tensor $P_{m\lambda_m j\lambda_j n\lambda_n}^{a\lambda_a b\lambda_b}=f^B_{b\lambda_b j\lambda_j n^*\lambda_n}f^A_{m\lambda_m a\lambda_a j^*\lambda_j}$. In general, $A \otimes B$ is reducible and can be decomposed into a direct sum of irreducible $A$-$B$-bimodules $M_{1}\oplus \, M_2 \oplus \, M_{3}\oplus \dots$. 
A bimoudle morphism $\chi:M\to N$ satisfies the following condition:
\be
\T\left(\bket{\BimA}\right)=\bket{\BimB}.
\ee

\bibliographystyle{apsrev4-1}
\bibliography{StringNet}

\end{document}